\documentclass[aps,prb,reprint,amsmath,amssymb,notitlepage,citeautoscript,superscriptaddress,showpacs,showkeys]{revtex4-2}

\usepackage{bm,mathrsfs,dcolumn,graphicx,color}
\usepackage{amsmath}
\usepackage{graphicx}
\usepackage[T1]{fontenc}
\usepackage{hyphenat}
\usepackage{setspace}

\usepackage[normalem]{ulem}
\definecolor{darkerblue}{rgb}{0,0,0.75}
\definecolor{darkerred}{rgb}{0.8,0,0}
\usepackage[colorlinks=true,citecolor=darkerblue,linkcolor=darkerred]{hyperref}

\definecolor{ablue}{rgb}{0.1,0.35,0.75}
\definecolor{agreen}{rgb}{0,0.55,0.3}
\definecolor{ared}{rgb}{0.8,0,0}
\definecolor{abrown}{RGB}{160,82,45}
\definecolor{mm}{rgb}{0.5,0.05,0.5}
\definecolor{orange}{RGB}{255,102,0}

\usepackage{hyperref}

\usepackage{enumitem} 

\begin{document}

\title{Electron recoil effect in electrically tunable MoSe$_2$ monolayers}

\author{Jonas Zipfel*}
\affiliation{Department of Physics, University of Regensburg, Regensburg D-93053, Germany}
\affiliation{Molecular Foundry, Lawrence Berkeley National Laboratory, Berkeley, California 94720, USA}
\author{Koloman Wagner*}
\affiliation{Department of Physics, University of Regensburg, Regensburg D-93053, Germany}
\affiliation{Dresden Integrated Center for Applied Physics and Photonic Materials (IAPP) and Würzburg-Dresden Cluster of Excellence ct.qmat, Technische Universität Dresden, 01062 Dresden, Germany}
\author{Marina A. Semina}
\affiliation{Ioffe Institute, Saint Petersburg, Russian Federation}
\author{Jonas D. Ziegler}
\affiliation{Department of Physics, University of Regensburg, Regensburg D-93053, Germany}
\affiliation{Dresden Integrated Center for Applied Physics and Photonic Materials (IAPP) and Würzburg-Dresden Cluster of Excellence ct.qmat, Technische Universität Dresden, 01062 Dresden, Germany}
\author{Takashi Taniguchi}
\affiliation{International Center for Materials Nanoarchitectonics, 
National Institute for Materials Science,  1-1 Namiki, Tsukuba 305-0044, Japan}
\author{Kenji Watanabe}
\affiliation{Research Center for Functional Materials, 
National Institute for Materials Science, 1-1 Namiki, Tsukuba 305-0044, Japan}
\author{Mikhail M. Glazov}
\affiliation{Ioffe Institute, Saint Petersburg, Russian Federation}
\author{Alexey Chernikov}
\email{alexey.chernikov@ur.de}
\affiliation{Department of Physics, University of Regensburg, Regensburg D-93053, Germany}
\affiliation{Dresden Integrated Center for Applied Physics and Photonic Materials (IAPP) and Würzburg-Dresden Cluster of Excellence ct.qmat, Technische Universität Dresden, 01062 Dresden, Germany}

\begin{abstract}
Radiative recombination of excitons dressed by the interactions with free charge carriers often occurs under simultaneous excitation of either electrons or holes to unbound states.
This phenomenon, known as the electron recoil effect, manifests itself in pronounced, asymmetric spectral lineshapes of the resulting emission.
We study the electron recoil effect experimentally in electrically-tunable monolayer semiconductors and derive it theoretically using both trion and Fermi-polaron pictures.
Time-resolved analysis of the recoil lineshapes is employed to access transient, non-equilibrium states of the exciton-carrier complexes.
We demonstrate cooling of the initially overheated populations on the picosecond timescales and reveal the impact of lattice temperature and free carrier density.
Both thermally activated phonons and the presence of free charges are shown to accelerate equilibration.
Finally, we find strong correlations between relaxation times from recoil analysis and luminescence rise times, providing a consistent interpretation for the initial dynamics of trion/Fermi-polaron states. 
\end{abstract}
\maketitle

\section{Introduction}

Light-matter coupling of bound electron-hole pairs, known as excitons, plays a fundamental role in the optical properties of semiconductors\,\cite{Haug2009}.
The nature of this coupling, however, can change substantially when additional quasiparticles are introduced\,\cite{Klingshirn2007}.
Among frequently encountered cases of interacting many-particle systems are mixtures of excitons with free charge carriers.
It leads to the formation of composite excitonic states, commonly labeled as trions\,\cite{Lampert1958,Kheng1993} or Fermi-polarons in the low-density limit\,\cite{Koudinov2014,Sidler2016,Efimkin2017}.
When these complexes recombine radiatively, their center-of-mass momenta are almost entirely transferred to the free carriers that remain in the system.
Broadly known as \textit{electron recoil} effect for trions\,\cite{Esser2001}, it expands the number of optically accessible states and alters both spectral shape and the dynamics of the emission.
This behavior strongly contrasts recombination of single excitons that is typically restricted to near-zero momenta within the light-cone\,\cite{Ivchenko2005}. 

Experimentally, electron recoil has been demonstrated in two-dimensional nanosystems such as quantum wells\,\cite{Esser2001,Esser2000a} and, more recently, in monolayers of transition-metal dichalcogenides (TMDCs)\,\cite{Ross2013}.
The latter are particularly well suited to explore exciton-electron phenomena due to strong Coulomb interactions\,\cite{Yu2015,Wang2018} that stabilize excitonic complexes even at elevated temperatures and high carrier densities. 
Interestingly, while the spectral asymmetry due to recoil is frequently observed in TMDCs\,\cite{Ross2013, Christopher2017, Lyons2019, Zhumagulov2020, Park2021}, it has been studied exclusively under steady-state conditions so far.
Time-resolved detection of the recoil flanks, however, would provide a direct access to non-equilibrium distributions and trion cooling dynamics.
These are typically challenging to access, often requiring advanced experimental approaches in combination with microscopic theory, as it has been recently demonstrated for the case of free electrons\,\cite{Venanzi2021}.
Accurate examination of the recoil effect should further profit from recent advances in material preparation using state-of-the-art encapsulation in hBN. 
In addition, recent analysis highlighted an alternative description of the excitons interacting with free carriers as attractive and repulsive Fermi-polarons\,\cite{Koudinov2014,Sidler2016,Efimkin2017,Chang2018,Fey2020,Cotlet2020,Glazov2020}.
A consistent theoretical description of the electron-assisted recombination of dressed exciton states in both trion and Fermi polaron pictures is thus highly desirable\,\footnote{Please also see a recent preprint discussing spectral lineshape asymmetry for Fermi polarons: T. Wasak et al. arXiv:2103.14040}.

Here, we address the above by studying electron recoil effect in two-dimensional TMDCs from both experimental and theoretical perspectives. 
Optical studies of field-effect transistors based on high-quality hBN-encapsulated MoSe$_2$ monolayers with thin-layer graphite gates are combined with analytical theory of trions and Fermi-polarons. 
It enables us to develop a consistent physical description of the interactions of excitons with resident charge carriers.
The manuscript is organized as follows: in Sect.\,\ref{exp-details} we summarize the employed techniques of material preparation and time-resolved microscopy. 
Section\,\ref{recoil-exp} provides an overview of the main experimental observables and their quantitative analysis within a phenomenological approach.
The theory of the recoil effect is then discussed in Sect.\,\ref{recoil-theor} justifying the description of the experimental data. 
Implications for light-matter coupling are derived in both trion and Fermi-polaron pictures.
Time-resolved measurements of the recoil effect are presented in Sect.\,\ref{cooling-temp}.
They provide access to the transient excess temperatures of trion/Fermi-polaron complexes and reveal accelerated cooling at elevated lattice temperatures on picosecond time-scales.
Similar behavior is observed for increased free carrier densities that facilitate relaxation, as discussed in Sect.\,\ref{cooling-density}.
The main results and conclusions are then summarized in Sec. \ref{conclusion}.

\section{Experimental details}
\label{exp-details}
The field-effect device under study was fabricated by mechanical exfoliation and dry-stamping\,\cite{Castellanos-Gomez2014} of bulk crystals on SiO$_2$/Si substrates with prepatterned gold electrodes.
A MoSe$_2$ monolayer was encapsulated between 10's of nm thick hBN layers.
Exfoliated thin-layer graphite flakes were used for top- and bottom-gates as well as to contact MoSe$_2$.
The quality of encapsulation was confirmed by observing spectrally narrow luminescence across sufficiently large areas of many $\mu$m$^2$ on the sample.
The device was fixed onto an electrically-contacted sample holder and placed into a microscopy cryostat.
It was cooled down to heat-sink temperatures between 5 and 50\,K.
Free charge carrier densities were tuned by changing the gate voltage and calibrated by evaluating relative spectral shifts between neutral excitons and trions (repulsive and attractive Fermi polarons, respectively), see Appendix\,\ref{app:Doping}.
Their values were varied across the range between $4\times10^{10}$ and $2\times10^{12}$\,cm$^{-2}$, consistent with those estimated from  a parallel-plate capacitor model at higher voltages.

For optical measurements we used a 140\,fs pulsed Ti:sapphire laser with a repetition rate of 80\,MHz as excitation source.
The laser was tuned to a photon energy of 1.657\,eV, corresponding to near-resonant excitation into the high-energy flank of the ground-state exciton of MoSe$_2$ at 1.642\,eV with an effective absorption of a few percent.
The incident light was focused onto the sample by a 60x microscope objective to a spot of 1\,$\mu$m diameter.
The excitation energy density was set to 6\,$\mu$Jcm$^{-2}$, resulting in the estimated injected electron-hole pair-density per pulse of several 10$^{11}$\,cm$^{-2}$.
The emitted photoluminescence (PL) signal was dispersed in a spectrometer and detected by a picosecond streak camera in synchroscan, photon-counting mode. 
Additional details on sample preparation and experimental procedures are given in the supplementary of Ref.\,\onlinecite{Wagner2020}.

\section{Electron recoil in $\mbox{MoSe}_2$}
\label{recoil-exp}
\begin{figure*}[t]
	\centering
			\includegraphics[width=13.0 cm]{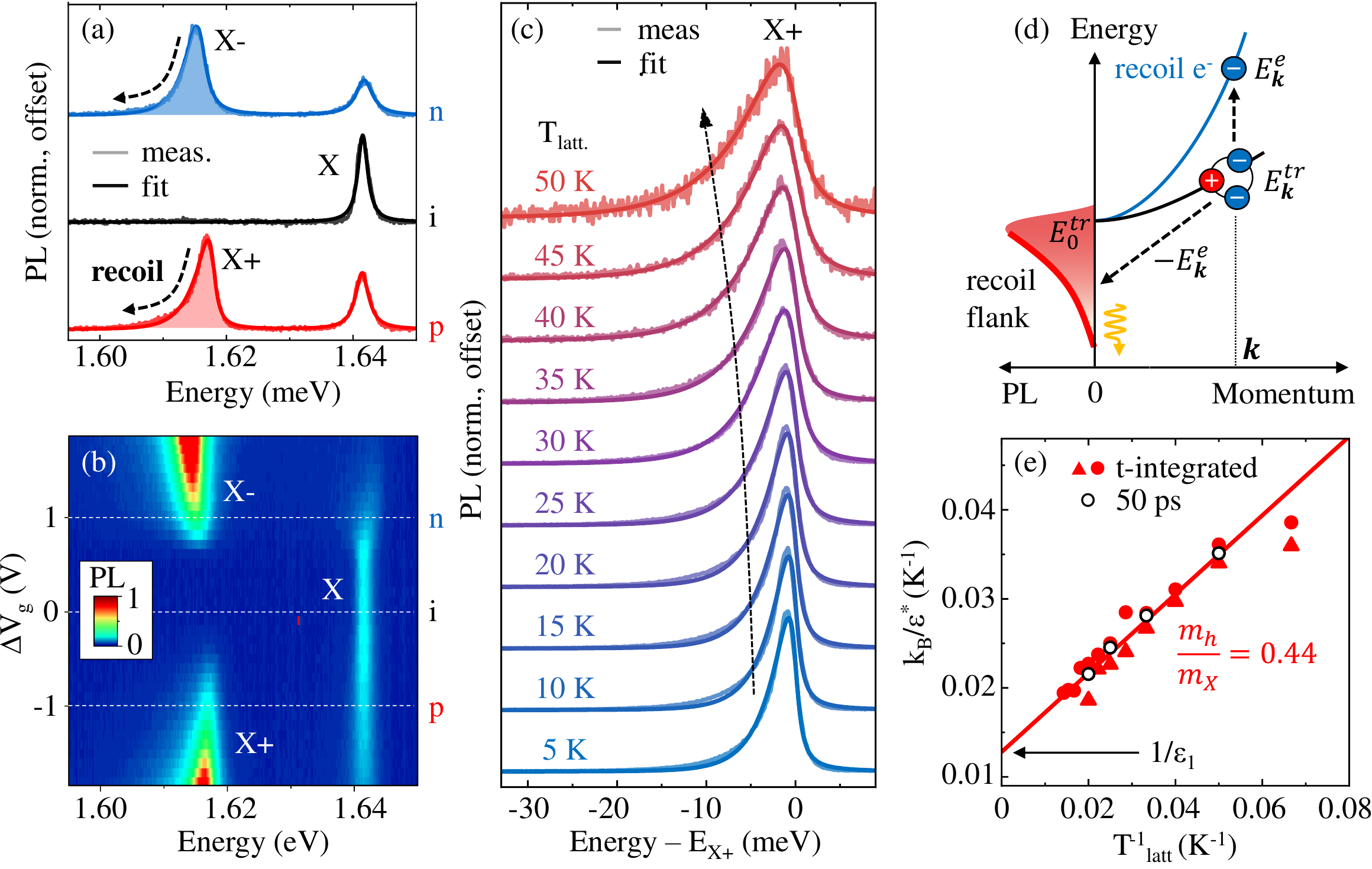}
		\caption{(a) Representative, time-integrated PL spectra of hBN-encapsulated MoSe$_2$ monolayer under electrical gating in the p-doped, neutral, and n-doped regimes at 5\,K. 
(b) False-color plot of the PL intensity as function of emission energy and gate voltage.
Dashed lines represent spectra presented in panel (a).  
The gate voltage of $\pm$1 V corresponds to the electron and hole density of about $1.6\times10^{11}$\,cm$^{2}$ and $2.9\times10^{11}$\,cm$^{2}$, respectively, estimated from the trion-exciton energy separation (at higher voltages the scaling is 2x10$^{11}$\,cm$^{2}$/V for holes and 4.8x10$^{11}$\,cm$^{2}$/V for electrons). 
(c) Time-integrated PL spectra at different lattice temperatures in the p-doped regime.
The hole density is set to $1\times10^{11}$\,cm$^{-2}$ for all temperatures. 
The spectra are presented as function of the emission energy relative to the zero-momentum trion peak energy $E_0^{tr}$ at each temperature. 
Fit functions are convolutions of a Lorentzian peak with an exponential low-energy flank as well as a Gaussian with a fixed linewidth of 1\,meV. 
(d) Schematic illustration of the electron recoil effect during trion recombination. 
(e) Exponential low-energy constant $\epsilon^*$ extracted from the fits in panel (c) (filled triangles), presented in the form of $k_B/\epsilon^*$ as function of the inverse lattice temperature for $T \geq 15$\,K. 
Filled circles correspond to a second measurement for temperatures up to 70\,K at a different position.
Fit of both data sets using Eq.\,\eqref{expon-fit} is shown in red.
Corresponding values obtained from time-resolved measurements at 50\,ps are shown by open circles.
}
	\label{fig1}
\end{figure*}

In this section we introduce the experimental observations of the recoil effect and the data analysis.
Time-integrated PL spectra of the gate-tunable MoSe$_2$ monolayer at 5\,K are presented in Fig.\,\ref{fig1}\,(a). 
The corresponding full set of PL measurements is shown in Fig.\,\ref{fig1}\,(b) as a function of gate voltage and emission energy.
Free carrier densities are estimated to be in the range of 10$^{11}$\,cm$^{2}$.
The data illustrate a typical response of MoSe$_2$\,\cite{Ross2013,Smolenski2019}, dominated by a single exciton resonance at charge neutrality (i).
Upon doping, an additional peak emerges that is shifted to lower energies by 27\,meV in the electron-doped regime (n) and by a slightly smaller value of 25\,meV in the case of hole-doping (p), consistent with the literature\,\cite{Shepard2017,Smolenski2019}.
It stems from the radiative recombination of a three-particle state known as trion\,\cite{Lampert1958,Kheng1993}, involving an exciton that binds an additional electron or a hole in an atom-like picture\,\cite{Rau1996}.
These quasiparticles can be also understood as attractive Fermi-polarons\,\cite{Ganchev2015,Sidler2016,Efimkin2017}, -- excitons dressed by a cloud of free carriers.
We note, that the two descriptions are conceptually equivalent at sufficiently low charge densities\,\cite{Glazov2020} studied here, and thus use the term trion throughout most parts of the manuscript.

While the neutral excitons in MoSe$_2$ emit as a narrow, symmetric peak with about 2\,meV linewidth, the trion PL is asymmetric with a broader low-energy flank.
The asymmetry becomes increasingly pronounced at higher temperatures, as observed in time-integrated spectra presented in Fig.\,\ref{fig1}\,(c) for the hole-doped regime. 
Such lineshapes are characteristic for trion emission\,\cite{Esser2000a} and are the consequence of the energy and momentum conservation during radiative recombination. 
The process is schematically illustrated in Fig.\,\ref{fig1}\,(d) adopting the picture of a n-doped case (it applies equally for p-doping).
A trion with the energy {$E_{\bm k}^{tr}$} recombines by emitting a photon and leaving a free electron.
Due to the vanishing momentum of the photon almost the entire trion momentum $\bm k$ is transferred to the remaining electron.
As a consequence, the electron gains the kinetic energy ${E_{\bm k}^e} \propto \bm k^2$, motivating the labeling of the effect as electron recoil in the literature\,\cite{Esser2001, Ross2013}.
Energy conservation then requires the photon being emitted at a lower energy than that of the trion by $E_{\bm k}^e$.
It leads to an asymmetric emission peak of the trions with a low-energy flank following their thermal distribution.

Assuming Boltzmann statistics, the resulting PL lineshape $I_{PL}$ can be thus described by a convolution of a symmetric peak function $I_0$ and an exponential function on the low-energy side: 
\begin{equation}
I_{PL}(\epsilon)=I_0 (\epsilon)\otimes \exp({\epsilon/\epsilon^*})\Theta({-\epsilon}),
\label{fit}
\end{equation}
with $\epsilon = {\hbar\omega - E_{0}^{tr}}$ denoting the photon energy ($\hbar\omega$)  relative to that of the trions at rest ($E_{0}^{tr}$) and $\Theta(\epsilon)$ representing the Heaviside step function; $\otimes$ denotes the convolution.
For the symmetric component $I_0 (\epsilon)$ we use a convolution of a Lorentzian peak and a Gaussian with a fixed linewidth of 1\,meV to account for homogeneous and inhomogeneous broadening, respectively.
The value $\epsilon^*$ in the low-energy exponent combines two contributions\,\cite{Esser2000a}:
\begin{equation}
\frac{1}{\epsilon^*}=\frac{1}{\epsilon_1}+\frac{1}{k_BT}\frac{m_e}{m_x}.
\label{expon-fit}
\end{equation}
The first term comes from an exponential approximation for the energy-dependent light-matter coupling  strength of the trions.
The constant $\epsilon_1$ is usually of similar magnitude as the trion binding energy\,\cite{Esser2000a}.
The second term accounts for electron recoil and depends on the trion temperature $T$ as well as the ratio of electron ($m_e$) and exciton ($m_x$) total translational masses.
The origin of Eq.\,\eqref{expon-fit} is discussed in detail in Sect.\,\ref{recoil-theor}. 

The model described by Eq.\,\eqref{fit} is used to fit time-integrated experimental spectra presented in Figs.\,\ref{fig1}\,(a) and (c). 
Setting symmetric broadening and the low-energy exponential $\epsilon^*$ as free parameters we obtain an overall good match to the measured lineshapes with only small deviations.
The extracted values for $\epsilon^*$ are presented in Fig.\,\ref{fig1}\,(e) in the form of $k_B/\epsilon^*$ as function of the inverse lattice temperature for $T >$10\,K.
The plot thus directly represents the relation in Eq.\,\eqref{expon-fit} with the slope being proportional to the mass ratio $m_h/m_x$ and the y-axis offset corresponding to $1/\epsilon_1$.
Only at the lowest temperatures of $T\leq$15\,K we find deviations from the linear behavior of $k_B/\epsilon^*$ as function of $T^{-1}$.
These mainly originate from the increased influence of overheated trion populations, as discussed further below in Sect.\,\ref{cooling-temp} and in Appendix\,\ref{app:Density}.

From the analysis presented in Fig.\,\ref{fig1}\,(e) we obtain the parameters of $m_h/m_x=0.44$ and $\epsilon_1=7$\,meV.
Both values are reasonable in view of the nearly equal electron and hole masses in the monolayer MoSe$_2$ with $m_x=m_e+m_h\approx 2m_h$ as well as the trion binding energies of a few 10's of meV.
The presented quantitative description of the temperature-dependent spectral asymmetry of the trion PL strongly supports the interpretation of this observation as the recoil effect.
It provides a solid basis for using time-resolved measurements of the recoil flanks to access transient trion temperatures and cooling dynamics presented in Sect.\,\ref{cooling-temp} and Sect.\ref{cooling-density}.
In the following Sect.\,\ref{recoil-theor} we start with the theoretical discussion of the electron recoil to outline the consequences for light-matter coupling and rationalize the applied lineshape analysis.

\section{Theory of electron recoil}
\label{recoil-theor}

In this section we derive the analytical expressions describing the PL spectra and recoil effect in a two-dimensional system. We use both trion and Fermi-polaron approaches and demonstrate that at sufficiently low densities of resident electrons they give essentially the same result, as it is the case for their spectral characteristics\,\cite{Glazov2020c}. We recall that the trion picture\,\cite{Lampert1958,Kheng1993} is a few-body approach that considers an exciton, a bound state of an electron and a hole, and a trion, a bound state of two electrons and a hole (or, in the case of p-type structure, of two holes and an electron). By contrast, in the Fermi-polaron approach we consider the exciton as a rigid particle immersed in the  Fermi-sea of electrons (or holes)\,\cite{Suris2001,suris:correlation,PhysRevA.85.021602,Koudinov2014,Sidler2016,Efimkin2017,Cotlet2019}.
The exciton is then ``dressed'' by the Fermi-sea excitations resulting in the formation of the attractive and repulsive Fermi-polaron states.

\subsection{Trion picture}\label{subsec:trion}

It is instructive to start the analysis of the photoluminescence spectra in terms of trions providing a convenient  physical picture of the effect, as illustrated in Fig.~\ref{fig1}(d).
In the trion language the rate of the photon emission with the frequency $\hbar\omega$ by the trion gas can be expressed using the Fermi golden rule in the form
\begin{equation}
\label{tr:FGR}
W^{tr}(\omega) = \frac{2\pi}{\hbar} \sum_{\bm k} f_{\bm k} |M^{tr,opt}(\bm k)|^2 \delta(\hbar\omega +E^e_{\bm k} - E^{tr}_{\bm k}).
\end{equation}
Here $E_{\bm k}^e= \hbar^2 k^2/2m_e$ and $E_{\bm k}^{tr}=E_0^{tr} + \hbar^2k^2/2m_{tr}$ are the electron and trion kinetic energies with $m_e$ and $m_{tr} = m_x+m_e$ being the electron and trion effective masses,
\begin{equation}
\label{f:tr}
f_{\bm k} = \mathcal N\exp{\left(- \frac{\hbar^2 k^2}{2m_{tr} k_B T_*} \right)}
\end{equation}
is the trion distribution function in the Boltzmann approximation.
Here, $T_*$ is the effective temperature of the trion gas which can, in the general case, deviate from that of the lattice and electrons, $\mathcal N$ is the normalization constant, and $M^{tr,opt}_{\bm k}$ is the matrix element of the trion radiative decay~\cite{Esser2000a,Glazov2020c}
\footnote{Here we correct a typo in Eqs. (24), (26) of Ref. \cite{Glazov2020c} where the factor $m_x/m_{tr}$ was missing in the exponent.}
\begin{equation}
\label{M:opt}
M^{tr,opt}(\bm k) =\mathfrak M_r \int \varphi(0,\bm \rho) e^{-\mathrm i \bm k \bm \rho (m_x/m_{tr})} d\bm \rho.
\end{equation}
Here $\varphi(\bm \rho_1,\bm \rho_2)$ is the trion envelope wavefunction, $\bm \rho_{1,2}$ are the relative coordinates of the electrons with respect to the hole. In derivation of Eq.~\eqref{tr:FGR} we further assumed that the electron density $N_e$ is sufficiently low such that $\hbar^2 N_e/m_e \ll E_{b,tr}, k_B T_*$ where $E_{b,tr}$ is the trion binding energy. In this case one can neglect modifications of the trion state by the resident carriers~\cite{Glazov2020c}. It also allows us to neglect the occupancy of the final electron state after the trion recombination.

\begin{figure}[tb]
	\centering
			\includegraphics[width=7.5 cm]{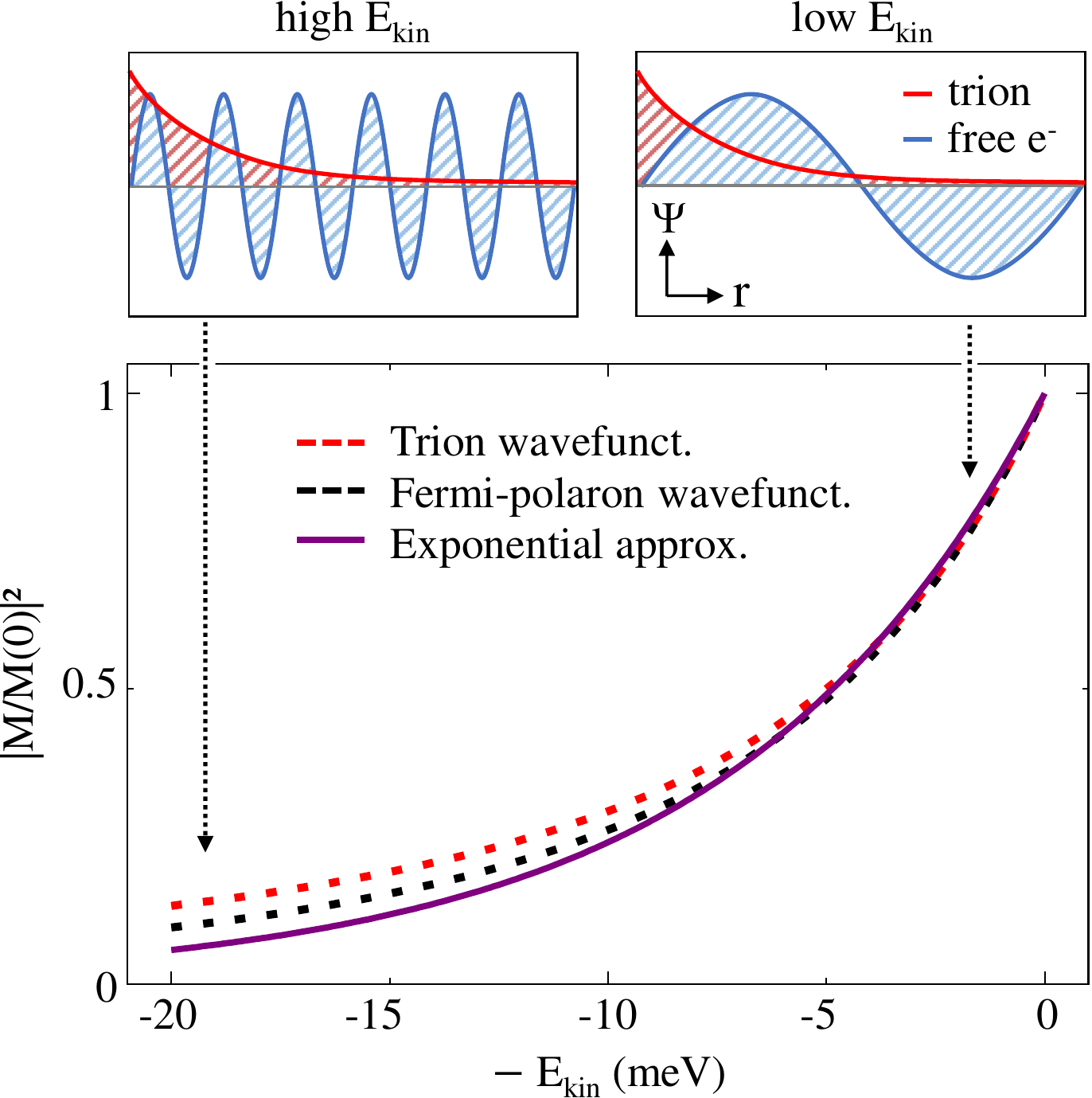}
		\caption{The top panels schematically illustrate overlap between trion/Fermi-polaron wavefunction with that of the final state of the free electron. 
Presented are two cases of high and low kinetic energies resulting in small and large overlap integrals, respectively.
The main panels shows calculated dipole matrix elements as function of the kinetic energy of the quasiparticle, using either trion or Fermi-polaron wavefunctions. Dashed black line is calculated after Eq.~\eqref{M:opt:trial} and dashed magenta line is calculated after Eq.~\eqref{M:fp:opt}.
Single-exponential approximation is presented by the solid line for direct comparison. 
For the trion we set $a_{tr}=3 a_x$ and obtain the binding energy of $17$~meV, close to the experimental values (we note that such deviations are common for variational models). 
For the Fermi-polaron approach we set $E_{b,tr}$ to the experimental value of $25$~meV and the exponential approximation corresponds to $\epsilon_1=7$~meV.
		}
	\label{fig2}
\end{figure} 

Altogether, Eq.~\eqref{tr:FGR} describes the radiative transition where the trion with the in-plane wavevector $\bm k$ recombines emitting a photon with negligible momentum and  leaving behind an electron with the wavevector $\bm k$.
As a consequence, the trion PL spectrum has a typical recoil low-energy wing in the form
\begin{multline}
\label{tr:FGR:1}
W^{tr}(\omega) \propto \exp{\left(\frac{\hbar\omega - E^{tr}_0}{k_B T_*} {\frac{m_e}{m_x}}\right)}\\
\times\left|M^{tr,opt}\left(\sqrt{\frac{2m_e m_{tr}(E^{tr}_0 - \hbar \omega)}{m_x\hbar^2}}\right)\right|^2\Theta(E^{tr}_0 - \hbar\omega).
\end{multline}

This expression agrees with Eqs.~\eqref{fit} and \eqref{expon-fit}, where, following Ref.~\cite{Esser2000a}, the energy dependence of $|M^{tr,opt}|^2$ has been approximated by the exponential function. 
To justify this assumption we use the trion envelope function in the simple form~\cite{Sergeev2001,Berkelbach2013,Courtade2017}
\begin{equation}
\label{trion:trial}
\psi(\bm \rho_1,\bm \rho_2) = e^{-\rho_1/a_x}e^{-\rho_2/a_{tr}} + e^{-\rho_2/a_x}e^{-\rho_1/a_{tr}},
\end{equation}
with $a_e$ and $a_{tr}$ being the effective ``exciton'' and ``trion'' Bohr radii, the matrix element is evaluated as
\begin{multline}
\label{M:opt:trial}
M^{tr,opt}(\bm k) =\frac{\mathfrak M_r}{\sqrt{a_{tr}^2a_x^2/8+2a_{tr}^4a_x^4/(a_{tr}+a_x)^4}} \\
\times \left(\frac{a_{tr}^2}{[1+a_{tr}^2 k^2 ({m_x}/m_{tr})^2]^{3/2}} \right.\\
\left.+ \frac{a_{x}^2}{[1+a_{x}^2 k^2 ({m_x}/m_{tr})^2]^{3/2}} \right).  
\end{multline}
Figure~\ref{fig2} shows the matrix element calculated after Eq.~\eqref{M:opt:trial} (dashed black  curve) and exponential approximation (solid magenta curve) as functions of the trion kinetic energy.
It demonstrates a good accuracy of the exponential approximation to describe the form of the function, especially in the energy range of the $\epsilon_1$ constant.

\subsection{Fermi-polaron picture}
\label{subsec:FP}

Within the Fermi-polaron approach the PL can be described as a process of the polaron recombination. After recombination, the Fermi-sea can either remain in the unperturbed state or acquire an excitation in the form of the Fermi-sea electron-hole pair (with the hole being an unoccupied state under the Fermi level) depending on the momentum of the polaron. As a result, a low-energy flank in the PL is formed.

\begin{figure}[t]
\includegraphics[width=\linewidth]{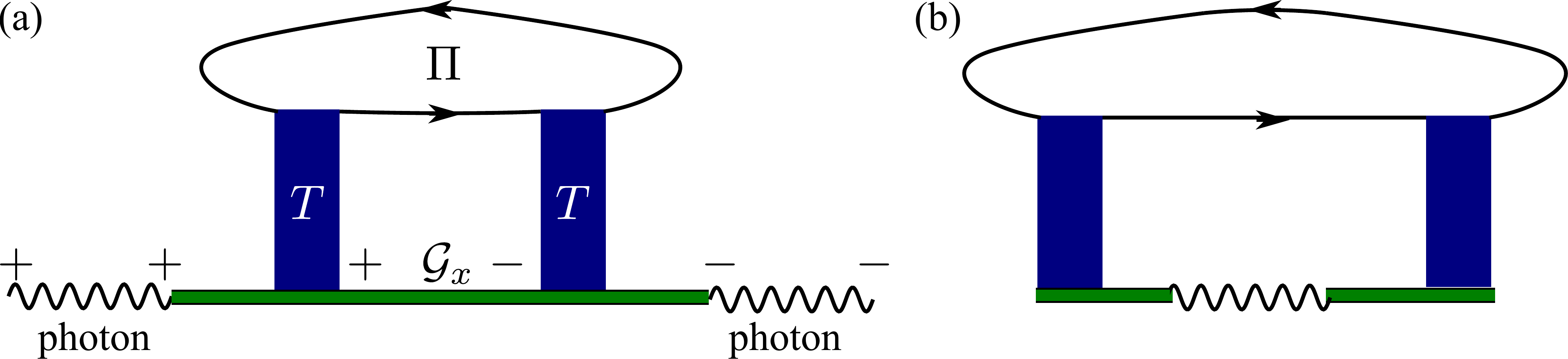}
\caption{Fermi-polaron interaction with photons. (a) The full diagram describing the Fermi-polaron PL in the Keldysh technique. (b) Self-energy responsible for the radiative decay. Wavy line shows the photon Greens function.
Exciton Greens function $\mathcal G_x$ renormalized by the Fermi-sea is shown by the thick green line.
Blue boxes show the electron-exciton scattering amplitude $T$, and $\Pi$ denotes the Fermi-sea polarization loop composed of the bare electron Greens functions (black lines with arrows). 
Signs $+$ and $-$ denote the components of the Greens functions in Keldysh technique.}\label{fig:K:PL}
\end{figure}

Fermi-polaron effects can be conveniently described using the diagram technique~\cite{suris:correlation,Efimkin2017,Cotlet2019,Glazov2020c}.
We consider the low electron density regime where, as above, $\hbar^2 N_e/m_e \ll E_{b,tr} \ll E_{b,x}$, evaluate the diagram in Fig.~\ref{fig:K:PL}(a) and derive the following expression for the PL spectrum {(using the non-equilibrium Keldysh technique, see Appendix~\ref{app:FPPL} for technical details)}: 
\begin{multline}
\label{PL:FP:Keld}
W^{FP}(\omega) = - \sum_{\bm k} Z_{\bm k} \mathrm \Im \{\Sigma_{PL}(\hbar\omega) \}
\\ = \frac{2\pi}{\hbar} \sum_{\bm k}  f_{\bm k} \frac{E_{b,tr}}{\mathcal D} \frac{|\mathfrak M_r|^2}{(\hbar\omega - E_0^x)^2}  \delta(\hbar\omega + E^e_{\bm k} - E_{\bm k}^{tr}).
\end{multline}
Here $\mathcal D={\mu_{tr}}/(2\pi\hbar^2) $ is the exciton-electron reduced density of states with $\mu_{tr} = m_e {m_x}/(m_e+{m_x})$ {and} $Z_{\bm k}\approx N_e /(\mathcal D E_{b,tr})$ is the renormalization factor.
The explicit expressions for the self-energy $\Sigma_{PL}(\hbar\omega)$ obtained via the Fermi-sea polarization loop $\Pi(\hbar\Omega, \bm k)$, scattering amplitude $T(\varepsilon,\bm k)$, and Greens functions are presented in the Appendix~\ref{app:FPPL}. Summing over the $\bm k$ we arrive at
\begin{equation}
\label{PL:FP:Keld:1}
W^{FP}(\omega) \propto  \exp{\left(\frac{\hbar\omega - E^{tr}_0}{k_B T_*}\frac{m_e}{m_x}\right)} |M^{fp, opt}|^2\Theta(E^{tr}_0 - \hbar\omega).
\end{equation}
The factor 
\begin{equation}
\label{M:fp:opt}
|M^{fp, opt}|^2 =  \frac{\mathfrak M_r^2}{\left(\hbar\omega - E_0^x \right)^2} = \frac{1}{(E_{\bm k}^{tr} - E_{\bm k}^e - E_{0}^x)^2},
 \end{equation} can be directly associated, in the model of the short-range exciton-electron interaction, with the squared absolute value of the matrix element in Eqs.~\eqref{M:opt} and \eqref{tr:FGR:1}.

Consequently, Eqs.~\eqref{tr:FGR:1} for trions and \eqref{PL:FP:Keld:1} for attractive Fermi-polarons are essentially equivalent. 
To illustrate it quantitatively, we show $|M^{fp, opt}|^2$ calculated after Eq.~\eqref{M:fp:opt} in Fig.~\ref{fig2} by dashed magenta curve. 
The result of the Fermi-polaron model is thus  in good agreement with the optical matrix element found with the trion wavefunction in Sec.~\ref{subsec:FP} and can also be quite accurately approximated by the exponential function.

To make our description of the trion/attractive Fermi-polaron PL complete, we also analyze the radiative decay rate of the trion (attractive Fermi polaron) ensemble.  To that end, we calculate the self-energy of the attractive polaron related to the light-matter coupling, Fig.~\ref{fig:K:PL}(b), see Appendix~\ref{app:rad} for details:
\begin{equation}
\label{sigma:k:phot:1}
\Sigma_{phot}(\varepsilon, \bm k) 
=
-\mathrm i \pi \frac{|\mathfrak M_r|^2}{(\varepsilon - E_{\bm k}^e - E_0^x)^2} N_e T^2(\varepsilon,\bm k) \mathcal D_{phot},
\end{equation} 
where $\mathcal D_{phot}$ is the density of photon states with the frequency $\omega=\varepsilon/\hbar$.
Using the Dyson equation $\mathbb G_x = \mathcal G + \mathcal G \Sigma \mathbb G$ we obtain the radiative damping rate of the Fermi-polaron with the energy $\varepsilon = E^{tr}_{\bm k} - N_e/\mathcal D$ in the form
\begin{equation}
\label{gamma:phot:diag}
\gamma_{r,\bm k} =  \frac{\pi}{\hbar} E_{b,tr}\frac{\mathcal D_{phot}}{\mathcal D}\frac{|\mathfrak M_r|^2}{(\varepsilon - E_{\bm k}^e - E_0^x)^2}.
\end{equation}
In derivation of Eqs.~\eqref{sigma:k:phot:1} and \eqref{gamma:phot:diag} we disregarded polarization-dependent factors, cf. Ref.~\cite{Esser2000a} which do not significantly affect the radiative decay rate of the quasiparticles ensemble. Equation~\eqref{gamma:phot:diag} can be recast as
\begin{equation}
\label{gamma:phot:diag:1}
\gamma_{r,\bm k} =  C (q_{phot} a_{tr})^2 \gamma_x\frac{E_{b,tr}}{(\hbar\omega - E_{\bm k}^e - E_0^x)^2},
\end{equation}
where $C\sim 1$ is a numerical constant, $\gamma_x$ is the light-cone exciton radiative decay rate, and   $q_{phot}$ is the photon wavevector at the frequency $\omega$.
Equations~\eqref{sigma:k:phot:1} and \eqref{gamma:phot:diag} thus take into account the radiative decay processes accompanied by the recoil effect.

However, for the Fermi-polarions with $k<q_{phot}$, \textit{direct} recombination is also possible.
In this case, the ``exciton'' component of the polaron has a small momentum corresponding to the light-cone and can decay radiatively without the necessity to transfer its momentum to the remaining electron. Corresponding contribution to the decay rate is directly proportional to the Fermi-polaron oscillator strength~\cite{Glazov2020c,imamoglu2020excitonpolarons} and given by (with polarization-dependent factors being omitted):
\begin{equation}
\label{rad:direct}
\gamma_{phot,\bm k} = Z_{\bm k} \gamma_x =  \frac{N_e}{\mathcal D E_{b,tr}} \gamma_x, \quad k<q_{phot}.
\end{equation}

The resulting \textit{total} radiative recombination rate is then obtained for the thermalized ensemble of the exciton-polarons  by averaging Eqs.~\eqref{gamma:phot:diag:1} and \eqref{rad:direct} with the quasiparticle distribution function $f_{\bm k}$, Eq.~\eqref{f:tr}. 
Omitting numerical coefficients, the resulting total rate can be estimated as:
\begin{equation}
\label{rad:total}
\gamma_{tot}  \sim \left[\frac{E_F}{k_B T_*} + 1 \right](q_{phot} a_{tr})^2  \gamma_x.
\end{equation}
This expression illustrates  the  interplay of two distinct Fermi-polaron decay mechanisms: At sufficiently low doping densities and elevated temperatures of the trion/Fermi-polaron ensemble, where $E_F \ll k_B T_*$, the radiative decay via the recoil effect is more important.
This is the regime relevant for our experiments. 

In  contrast, for the high doping/low temperature conditions of Fermi-polaron ensemble, $E_F \gg k_B T_*$, the decay via the exciton component is dominant and takes place via the Fermi-polarons within the light-cone.
For the charge carrier masses of MoSe$_2$ on the order of $0.5\,m_0$  and trion temperatures of 10's of K, this high doping regime should roughly correspond to carrier densities in the $10^{12}$\,cm$^{-2}$ range.
Then, one would need to modify the spectral lineshape analysis of Eq.\eqref{fit} by introducing an additional symmetric peak to account for direct recombination of Fermi-polarons.

Note that for the repulsive Fermi-polarons (excitons) the lineshape asymmetry is expected to be much weaker.
Indeed, the repulsive polaron recombination takes place mainly via its excitonic component (which is significant as compared with the attractive polaron) making the effect of the recoil smaller. 
The PL asymmetry may arise in this case mainly due to the exciton-phonon interaction instead, see Ref.~\cite{Christiansen2017,shree2018exciton}.

\section{Trion/Fermi-polaron cooling}
\label{cooling-temp}

In this section we discuss time-resolved analysis of the recoil effect as well as temperature-dependent cooling dynamics of the trions.
We briefly illustrate theoretical considerations of relaxation time-scales followed by presenting experimental results and discussion.

\subsection{Model}
\label{cooling-temp:model}

A theoretical description of the cooling effects can be performed within either the trion or Fermi-polaron approach. 
Provided that the resident electron density is low enough and the quasiparticles temperature $T^*$ is sufficiently large such that ${E_F} \ll k_B T^*$ both approaches yield the same results. Here we use the trion picture; the same result can be derived in the Fermi-polaron approach, see Appendix~\ref{app:cool}. 

In particular,  we consider cooling of trions due to emission of longitudinal acoustic phonons.
Their interaction occurs via the deformation-potential mechanism with the bare matrix element of the trion-phonon interaction represented in the form~\cite{gantmakher87}
\begin{equation}
\label{me}
\left|M_{\bm k\to \bm k'}^{tr,\pm \bm q}\right|^2 = {\delta_{\bm k, \bm k\mp \bm q}}\mathcal B_{tr}(q), 
\end{equation}
where the top (bottom) sign refers to the  phonon emission (absorption), the normalization area, as above, is set  to unity, and the $\mathcal B(q)$ is reduced rate. It reads (cf.~\cite{Kaasbjerg2012,shree2018exciton,Glazov2020})
\begin{equation}
\label{def:ac}
\mathcal B_{tr}(q) = \frac{\hbar}{2\varrho  s} \times
\begin{cases}
(2\Xi_c - \Xi_v)^2, \quad  \mbox{n-type,}\\
(2\Xi_v - \Xi_c)^2, \quad \mbox{p-type,}
\end{cases}
\end{equation}
where $\Xi_c$, $\Xi_v$ are the conduction and valence band deformation potentials, $\varrho$ is the mass density of the monolayer, $s$ is the longitudinal sound speed, and we consider both n- and p-type structures.
In derivation of Eq.~\eqref{def:ac} we disregarded the form-factor arising from the overlap of the phonon and trion wavefunctions assuming that $qa_{tr} \ll 1$.

\begin{figure*}[ht]
	\centering
			\includegraphics[width=13.0 cm]{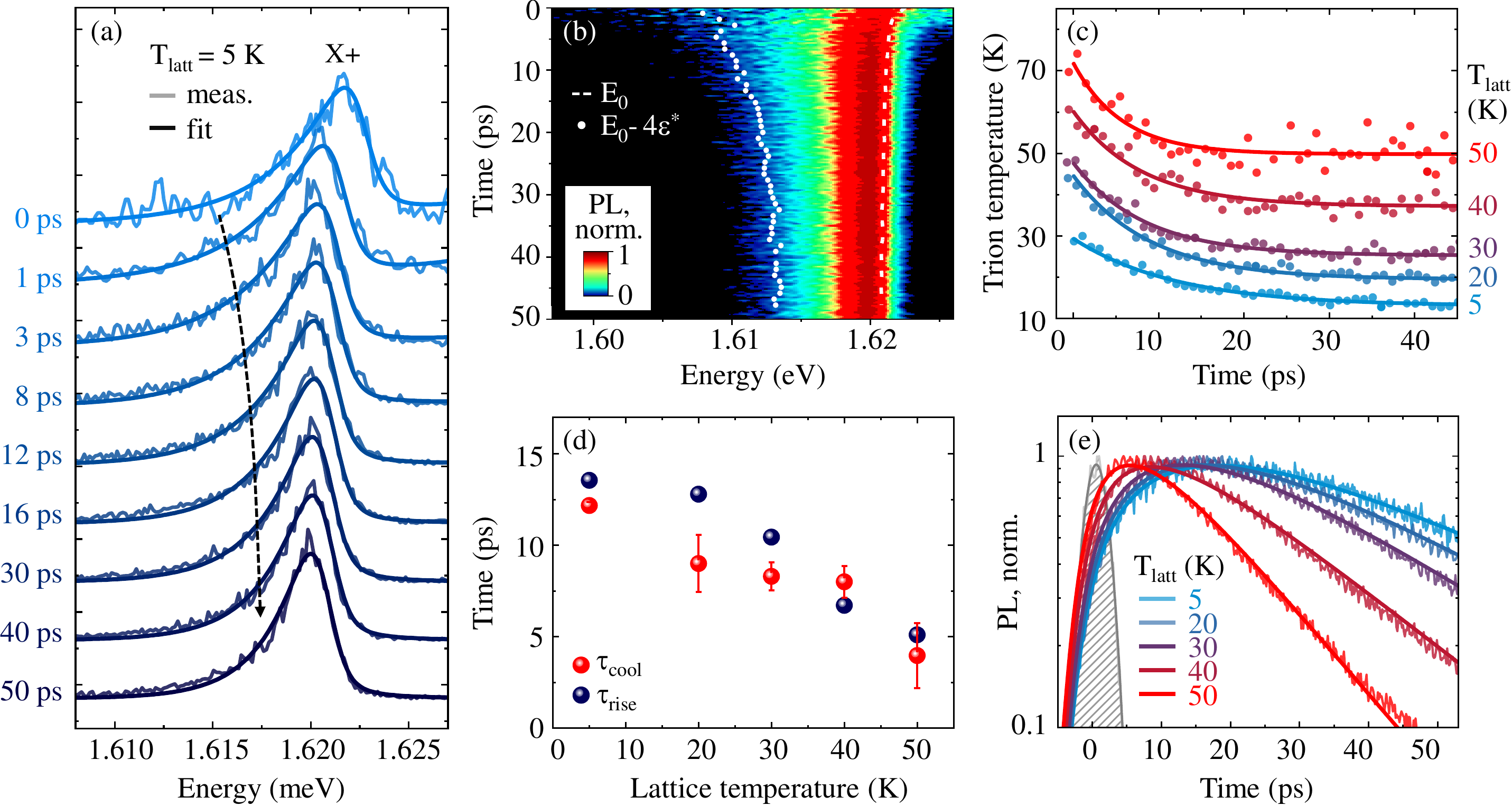}
		\caption{(a) Series of time-resolved PL spectra of the X+ resonance at 5 K for different times after near-resonant excitation of the neutral exciton at 1.663 eV.
Estimated pump density of the injected electron-hole pairs is $4.5\times 10^{11}$ cm$^{-2}$, considering absorption of 2\% at the pump energy. 
The density of the free holes is set to about $7\times 10^{10}$ cm$^{-2}$. 
(b) Corresponding streak camera image of the PL intensity in a false-color plot, as function of time and emission energy. 
Dotted and dashed lines represent the time-dependent exponent $\epsilon^*$ of the low-energy flank (x4) and the trion peak energy $E^{tr}_0$, respectively. 
They are obtained by an automated fit procedure of spectra in intervals of 1 ps using Eq.\,\eqref{fit}. 
(c) Extracted, transient temperatures of the trion measured for different lattice temperatures between 5 and 50\,K. 
(d) Characteristic 1/e cooling times of the trions as function of lattice temperature. 
Rise-time of the PL transients is presented for direct comparison. 
(e) Corresponding PL transients of the trion emission. Laser scattering signal representing instrument response function with 4\,ps width is shown by the gray area. 
An exponential rise-decay model was used for fitting, convoluted with a Gaussian to account for the instrument response.
}
	\label{fig3}
\end{figure*}

The energy loss rate by the non-degenerate trion gas characterized by the distribution function $f_{\bm k}$ reads
\begin{equation}
\label{Q:gen}
Q = \frac{1}{\sum_{\bm k} f_{\bm k}} \sum_{\bm k\bm q}  \hbar\omega^{ph}_{\bm q} \left[f_{\bm k} W_{\bm k\to \bm k-\bm q}^{+ \bm q} - f_{\bm k} W^{-\bm q}_{\bm k\to \bm k +\bm q}\right],
\end{equation}
where $ \hbar\omega^{ph}_{\bm q} = \hbar s q$ is the phonon dispersion,
\begin{equation}
\label{rate:trions}
W_{\bm k\to \bm k'}^{\pm \bm q} = \frac{2\pi}{\hbar} \left|M_{\bm k\to \bm k'}^{tr,\pm \bm q}\right|^2 \left(n^{ph}_{\bm q} +\frac{1}{2} \pm \frac{1}{2}\right) \delta(E^{tr}_{\bm k} - E^{tr}_{\bm k'} \pm \hbar\omega^{ph}_{\bm q}),
\end{equation}
$n^{ph}_{\bm q}$ is the phonon distribution function. Making the replacement $\bm k\to \bm k-\bm q$ in the second term in Eq.~\eqref{Q:gen} and using Eq.~\eqref{me} we arrive at 
\begin{multline}
\label{Q:1}
Q = \frac{1}{\int d\bm k f_{\bm k}} \int d\bm k \int \frac{d\bm q}{2\pi \hbar} \hbar\omega_{\bm q} \mathcal B(q) \delta (E^{tr}_{\bm k} - E^{tr}_{\bm k- \bm q} - \hbar\omega^{ph}_{\bm q}) \\
\times
\left[f_{\bm k}  (1+n^{ph}_{\bm q})- f_{\bm k-\bm q} n^{ph}_{\bm q}\right].
\end{multline}
Making use of the explicit form of the distribution function Eq.~\eqref{f:tr} and assuming that phonons are characterized by the lattice temperature $T$, Eq.~\eqref{Q:1} can be recast as
\begin{equation}
\label{Q:fin}
Q = \frac{k_B (T_*-T)}{\tau_{tr,T}},
\end{equation}
where the cooling time $\tau_T$ is defined as
\begin{equation}
\label{tau:T:1}
\frac{1}{\tau_{tr,T}} =   \frac{2m_{tr}^2 }{\hbar^3\varrho}
\times
\begin{cases}
(2\Xi_c - \Xi_v)^2, \quad  \mbox{n-type,}\\
(2\Xi_v - \Xi_c)^2, \quad \mbox{p-type.}
\end{cases}
\end{equation}
Note, that the cooling time of excitons is described by the same expression with the replacements $m_{tr} \to M_X$ and $(2\Xi_c - \Xi_v)^2\to (\Xi_c - \Xi_v)^{2}$~\cite{Glazov2020}.
We should thus expect the trions to cool at least as efficiently as the excitons and much faster than free charge carriers due to the larger mass.

\subsection{Experiment}
\label{cooling-temp:exp}

In the experiment we focus on the p-doped regime due to the overall narrower linewidths that allow for a more accurate quantitative analysis (similar observations for n-doping are presented in the Appendix\,\ref{app:nDoping}). 
Fig.\,\ref{fig3}\,(a) presents PL spectra of the trion emission obtained at different times after near-resonant excitation of the neutral exciton with the photon energy of 1.663\,eV.
Corresponding streak camera image of the emission is shown in Fig.\,\ref{fig3}\,(b).
The spectra are normalized at each time step for better lineshape comparison.

Directly after the excitation, the PL exhibits a broad low-energy recoil flank that subsequently narrows at later times.
It shows that trions are initially overheated with comparatively high kinetic energies.
Over time, they cool down towards thermal equilibrium determined by the lattice temperature.
In addition, the emission maximum shifts slightly in energy during the first few picoseconds, following the shift of the neutral exciton resonance.
The latter is a common observation upon optical excitation of semiconductors and is often related to the interplay of bandgap and exciton binding energy renormalization at finite excitation densities\,\cite{Haug2009,Steinhoff2014,Schmidt2016}.
   
For quantitative analysis of the equilibration dynamics we follow the procedure outlined in Sect.\,\ref{recoil-exp}. 
We use the model from Eq.\,\eqref{fit} to fit the spectra at each time step.
The resulting fit curves are shown in Fig.\,\ref{fig3}\,(a) and the corresponding low-energy exponents $\epsilon^*$, multiplied by factor of 4 for better illustration, are indicated in Fig.\,\ref{fig3}\,(b).
From $\epsilon^*$ we extract the effective, transient temperature of the trions by using Eq.\,\eqref{expon-fit}.
We fix the exponential parameter $\epsilon_1$ representing light-matter coupling dependence on kinetic energy to the previously determined value of 7\,meV, see Fig.\,\ref{fig1}\,(e).
The presented time-resolved measurements of the electron recoil and their analysis are then repeated for a series of lattice temperatures up to 50\,K.

The obtained effective temperatures of the trions are presented in Fig.\,\ref{fig3}\,(c) as function of time after the excitation.
For all studied lattice temperatures, the initial, non-equilibrium trion temperature is higher than that of the lattice by about 20\,K.
This value does not depend strongly on the excess energy of the pump laser, as we obtain a similar excess temperature also for non-resonant excitation conditions (see Appendix\,\ref{app:OffRes}).
It is thus likely to represent the residual excess energy of the trions after their formation on a picosecond timescale.
At later times the effective trion temperature decreases and reaches values that are indeed very close to the respective lattice temperatures.
Only at 5\,K the trions do not seem to cool down sufficiently, equilibrating above 10\,K.
The latter stems likely from additional heating of the system by the laser illumination, discussed in the Appendix\,\ref{app:Density}.

As illustrated in Fig.\,\ref{fig3}\,(c), the trion temperature decays over time and is well described by a single-exponential law.
Corresponding cooling times are summarized in Fig.\,\ref{fig3}\,(d) and are on the order of 10 to 15\,ps at the lowest studied temperatures.
These values are consistent with the expectations from the phonon-assisted cooling involving long-range acoustic phonon modes.
They roughly correspond to the relaxation time-scales previously determined for excitons in TMDCs\,\cite{Fang2019, Rosati2020} at liquid helium temperature in agreement with Eq.~\eqref{tau:T:1} which predicts about the same order of magnitude for the exciton and trion cooling times.
In contrast, trion cooling seems indeed to be faster than that of free electrons in MoSe$_2$, recently determined to occur within 70\,ps at 5\,K\,\cite{Venanzi2021}.
Similar to electrons, however, trions are found to equilibrate more rapidly at higher temperatures indicating additional contributions from thermally activated optical and zone-edge phonons\,\cite{Selig2016, Venanzi2021}.

The independent access to cooling times from transient recoil analysis further allows for a consistent interpretation of the PL intensity dynamics of the trions, presented in Fig.\,\ref{fig3}\,(e).
In general, it is \textit{a priori} unclear whether the PL rise time, for example, corresponds to relaxation or recombination, being determined by the faster one of the two processes.
For bright excitons in MoSe$_2$ it was indeed shown that a much shorter radiative lifetime governs the rise time\,\cite{Fang2019} and the slower relaxation time on the order of 20\,ps accounts for the PL decay.
In the present case of the trions, however, the extracted rise time closely follows the values of the cooling time, as illustrated in Fig.\,\ref{fig3}\,(d).
This result seems reasonable in view of the low doping densities $<10^{11}$\,cm$^{-2}$ used in our study.
In this regime, the transfer of the oscillator strength from the neutral exciton to the trion/attractive Fermi-polaron is rather small and should thus lead to comparatively long radiative lifetimes.
Consequently, the recombination of the trions, whether radiative or non-radiative, then determines the subsequent decay of the PL.

\section{Impact of free carriers on cooling}
\label{cooling-density}

In this section we address the trion cooling mediated by free carriers that can contribute or even dominate relaxation dynamics at finite densities in addition to phonons. 
Experimentally, we elucidate the role of free carriers for trion cooling from time-resolved analysis of the recoil effect as function of charge carrier density in the p-doped regime.

\begin{figure}[t]
	\centering
			\includegraphics[width=8.4 cm]{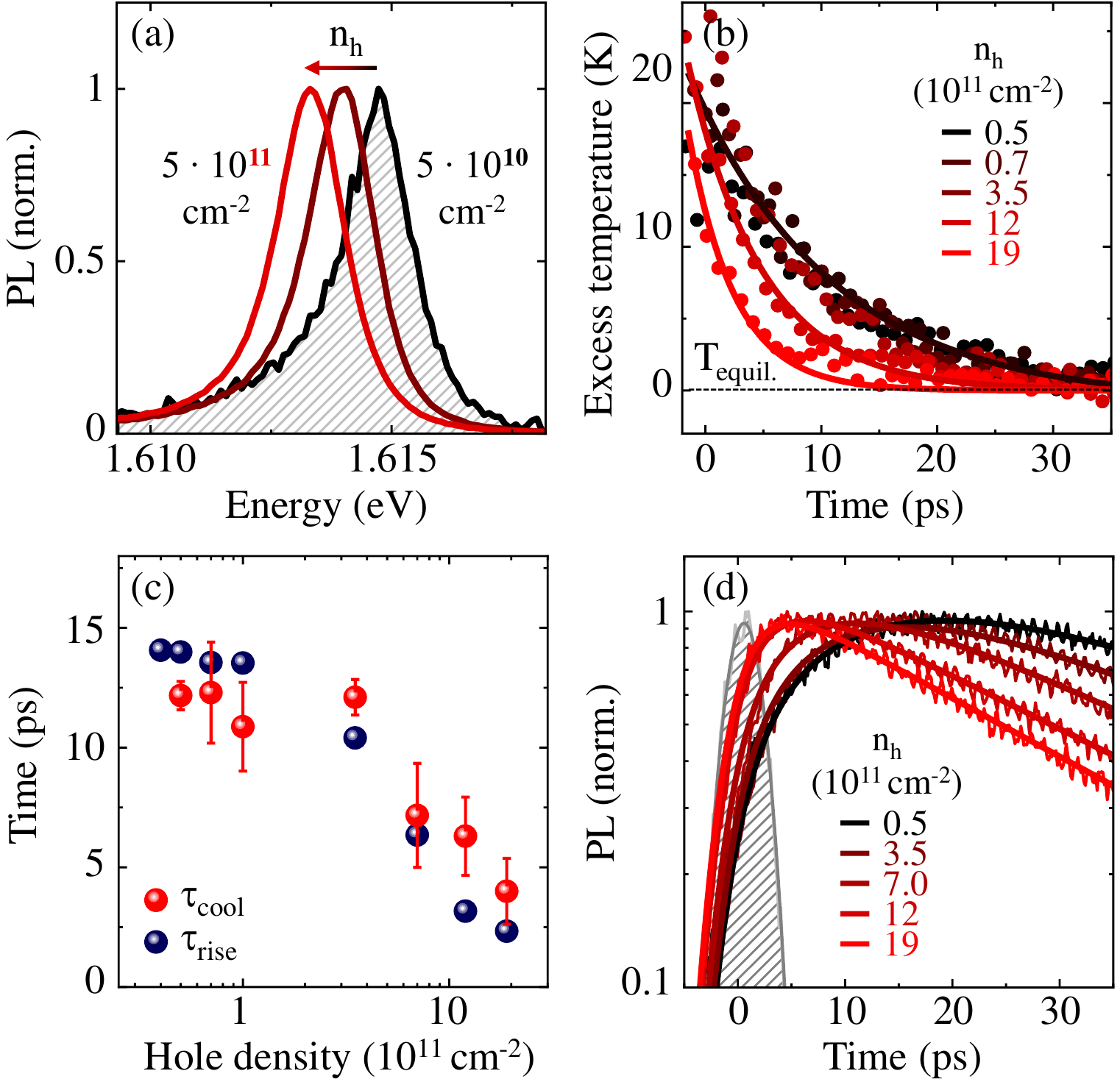}
		\caption{(a) Representative, time-integrated PL spectra of X+ at different doping densities up to $5\times 10^{11}$ cm$^{-2}$ (using the hole mass of 0.6\,m$_0$, 1 meV energy shift corresponds to change of doping by $2.5\times 10^{11}$ cm$^{-2}$).
		All experimental parameters are kept the same as for the measurements presented in Fig.\,\ref{fig3}: photon energy of 1.663 eV, pump density of 6 $\mu$J/cm$^{-2}$, estimated injected electron-hole density of $4.5\times 10^{11}$ cm$^{-2}$, and T\,=\,5\,K. 
		(b) Transient excess temperatures of the X$^+$ trion for different densities of the free holes with respect to the equilibrium temperature at 50\,ps. 
		(c) Extracted cooling time from the exponential fit as function of hole density. 
		Rise time of the trion PL is shown for comparison. 
		(d) Corresponding PL transients of the trion emission together with exponential rise-decay fits $\propto (\tau_{decay}-\tau_{rise})^{-1}[\exp(-t/\tau_{decay})-\exp(-t/\tau_{rise})]$.
}
	\label{fig4}
\end{figure}

Typical, time-integrated PL spectra of the X+ trion emission at 5\,K are presented in Fig.\,\ref{fig4}\,(a) for hole doping densities between $5\times 10^{10}$ cm$^{-2}$ and $5\times 10^{11}$ cm$^{-2}$.
While at the lowest doping density the spectral asymmetry from recoil effect is rather pronounced, the spectra acquire a more symmetric lineshape at higher doping densities. 
This observation is partially related to a slightly increased symmetric broadening due to additional density-induced scattering\,\cite{Efimkin2017,Carbone2020,Wagner2020} as well as potential contributions from direct recombination discussed in the context of Eqs.\,\eqref{rad:direct} and \eqref{rad:total}.
However, it is also associated with increasingly faster trion cooling that leads to a less pronounced asymmetry in the time-integrated response.

The doping-induced increase in the trion cooling rates is illustrated in Fig.\,\ref{fig4}\,(b) by evaluating the change of the excess temperature as function of time for a series of free hole densities.
Here, we used the same procedure as outlined in Sect.\,\ref{cooling-temp} for the time-resolved analysis of the recoil effect.
The resulting density-dependent cooling times obtained from single-exponential fitting of the temperature transients are summarized in Fig.\,\ref{fig4}\,(c).
In the low-density doping regime, below $3\times 10^{11}$ cm$^{-2}$, the trion cooling does not depend on the density of free carriers and is mediated by scattering with phonons.
At elevated hole densities, the cooling times decrease below 5\,ps at $2\times 10^{12}$ cm$^{-2}$.
The corresponding, hole-mediated cooling rate coefficient obtained from the inverse slope of the time constants is 0.042\,cm$^2$/s.
It means that at the free carrier density of $10^{11}$ cm$^{-2}$ the corresponding contribution to the trion cooling time is above 200\,ps and decreases to the range of 10's of ps for densities above $10^{12}$ cm$^{-2}$.
At these high densities and above relaxation of the trions is almost exclusively governed by scattering with free carriers.

Similar to the temperature-dependent analysis, we also find very similar cooling and PL rise times, as illustrated in Fig.\,\ref{fig4}\,(c).
Corresponding transients are plotted in Fig.\,\ref{fig4}\,(d) together with exponential rise-decay fits.
These results show that while direct radiative recombination of the trions / attractive Fermi-polarons is expected to occur at elevated doping levels, cooling processes largely determine the rise time over a broad range of free carrier densities.

\section{Conclusion}
\label{conclusion}

To summarize, we have studied electron recoil effect in charge-tunable MoSe$_2$ monolayers employing time-resolved photoluminescence spectroscopy (Sect.\,\ref{exp-details}) and analytical theory (Sect.\,\ref{recoil-theor}).
Characteristic spectral asymmetry of the bound states between excitons and free charge carriers was identified to originate from carrier-assisted radiative recombination: a photon is emitted and the free carrier system is left in an excited state (Sect.\,\ref{recoil-exp}).
Theoretically, this process was shown to occur using either a three-particle trion or a many-particle Fermi polaron description of the exciton-carrier mixtures (Sect.\,\ref{recoil-theor}).
In both formalisms the low-energy flank of the emission is determined by the thermal distribution of the quasiparticles and momentum-dependent light-matter coupling strength at sufficiently low carrier densities with $E_F<k_BT$.
The resulting spectral lineshapes of trion/Fermi-polaron luminescence was shown to provide an accurate quantitative description of temperature-dependent spectra (Sect.\,\ref{recoil-exp}).

Based on this analysis, time-resolved measurements of the recoil effect allowed for a direct evaluation of the transient trion temperature and relaxation dynamics (Sect.\,\ref{cooling-temp}).
Trions were shown to cool down on a timescale of about 15\,ps at $T=5K$, determined by scattering with long-wavelength acoustic phonons. 
The cooling process became increasingly faster at elevated temperatures, attributed to thermal activation of additional phonon modes.
Similarly, free carriers were found to mediate trion/Fermi-polaron cooling and dominate it at doping densities close to $10^{12}$ cm$^{-2}$ and above (Sect.\,\ref{cooling-density}).
Finally, we observed a close correspondence between relaxation timescales obtained from recoil analysis and rise times of the PL transients, providing a consistent interpretation of the PL dynamics.

Overall, direct experimental access to the non-equilibrium states of bound exciton-carrier complexes and their relaxation dynamics should be very useful for the broader community working on exciton-carrier mixtures in monolayer materials and their heterostructures.
Theoretically demonstrated agreement of trion and Fermi-polaron pictures for the appropriate description of the recoil effect in the low-density regime should be particularly helpful to motivate and support future work in this field.
Open questions include the interplay of direct and carrier-assisted recombination, radiative coupling in the high-density regime, as well as relaxation of higher-particle complexes such as neutral and charged biexcitons.

~\\
\section{Acknowledgments}
We thank Andrea Bergscheider, Martin Kroner, and Malte Selig for fruitful discussions as well as Christian B\"auml and Nicola Paradiso for their assistance with pre-patterned substrate preparation. 
Financial support by the DFG via Emmy Noether Initiative (CH 1672/1, Project-ID: 287022282), Walter Benjamin Programme (Project-ID: 462503440, J.Z.), SFB 1277 (project B05), as well as the W\"urzburg-Dresden Cluster of Excellence on Complexity and Topology in Quantum Matter ct.qmat (EXC 2147, Project-ID 390858490) is gratefully acknowledged.
K.Watanabe and T.T. acknowledge support from the Elemental Strategy Initiative conducted by the MEXT, Japan (Grant Number JPMXP0112101001) and  JSPS KAKENHI (Grant Numbers 19H05790, 20H00354 and 21H05233).
The development of the analytical theory by M.M.G. has been supported by RSF Project No. 19-12-00051; numerical calculations of the trion and Fermi-polaron states by M.A.S. have been supported by RSF Project No. 19-12-00273.
J. Z. and K. Wagner contributed equally to this manuscript.


\begin{thebibliography}{56}%
\makeatletter
\providecommand \@ifxundefined [1]{%
 \@ifx{#1\undefined}
}%
\providecommand \@ifnum [1]{%
 \ifnum #1\expandafter \@firstoftwo
 \else \expandafter \@secondoftwo
 \fi
}%
\providecommand \@ifx [1]{%
 \ifx #1\expandafter \@firstoftwo
 \else \expandafter \@secondoftwo
 \fi
}%
\providecommand \natexlab [1]{#1}%
\providecommand \enquote  [1]{``#1''}%
\providecommand \bibnamefont  [1]{#1}%
\providecommand \bibfnamefont [1]{#1}%
\providecommand \citenamefont [1]{#1}%
\providecommand \href@noop [0]{\@secondoftwo}%
\providecommand \href [0]{\begingroup \@sanitize@url \@href}%
\providecommand \@href[1]{\@@startlink{#1}\@@href}%
\providecommand \@@href[1]{\endgroup#1\@@endlink}%
\providecommand \@sanitize@url [0]{\catcode `\\12\catcode `\$12\catcode
  `\&12\catcode `\#12\catcode `\^12\catcode `\_12\catcode `\%12\relax}%
\providecommand \@@startlink[1]{}%
\providecommand \@@endlink[0]{}%
\providecommand \url  [0]{\begingroup\@sanitize@url \@url }%
\providecommand \@url [1]{\endgroup\@href {#1}{\urlprefix }}%
\providecommand \urlprefix  [0]{URL }%
\providecommand \Eprint [0]{\href }%
\providecommand \doibase [0]{https://doi.org/}%
\providecommand \selectlanguage [0]{\@gobble}%
\providecommand \bibinfo  [0]{\@secondoftwo}%
\providecommand \bibfield  [0]{\@secondoftwo}%
\providecommand \translation [1]{[#1]}%
\providecommand \BibitemOpen [0]{}%
\providecommand \bibitemStop [0]{}%
\providecommand \bibitemNoStop [0]{.\EOS\space}%
\providecommand \EOS [0]{\spacefactor3000\relax}%
\providecommand \BibitemShut  [1]{\csname bibitem#1\endcsname}%
\let\auto@bib@innerbib\@empty
\bibitem [{\citenamefont {Haug}\ and\ \citenamefont {Koch}(2009)}]{Haug2009}%
  \BibitemOpen
  \bibfield  {author} {\bibinfo {author} {\bibfnamefont {H.}~\bibnamefont
  {Haug}}\ and\ \bibinfo {author} {\bibfnamefont {S.~W.}\ \bibnamefont
  {Koch}},\ }\href@noop {} {\emph {\bibinfo {title} {{Quantum theory of the
  optical and electronic properties of semiconductors}}}},\ \bibinfo {edition}
  {5th}\ ed.\ (\bibinfo  {publisher} {World Scientific, Singapore},\ \bibinfo
  {year} {2009})\BibitemShut {NoStop}%
\bibitem [{\citenamefont {Klingshirn}(2007)}]{Klingshirn2007}%
  \BibitemOpen
  \bibfield  {author} {\bibinfo {author} {\bibfnamefont {C.}~\bibnamefont
  {Klingshirn}},\ }\href@noop {} {\emph {\bibinfo {title} {{Semiconductor
  Optics}}}},\ \bibinfo {edition} {3rd}\ ed.\ (\bibinfo  {publisher} {Springer,
  Berlin Heidelberg New York},\ \bibinfo {year} {2007})\BibitemShut {NoStop}%
\bibitem [{\citenamefont {Lampert}(1958)}]{Lampert1958}%
  \BibitemOpen
  \bibfield  {author} {\bibinfo {author} {\bibfnamefont {M.~A.}\ \bibnamefont
  {Lampert}},\ }\bibfield  {title} {\bibinfo {title} {{Mobile and Immobile
  Effective-Mass-Particle Complexes in Nonmetallic Solids}},\ }\href
  {https://doi.org/10.1103/PhysRevLett.1.450} {\bibfield  {journal} {\bibinfo
  {journal} {Phys. Rev. Lett.}\ }\textbf {\bibinfo {volume} {1}},\ \bibinfo
  {pages} {450} (\bibinfo {year} {1958})}\BibitemShut {NoStop}%
\bibitem [{\citenamefont {Kheng}\ \emph {et~al.}(1993)\citenamefont {Kheng},
  \citenamefont {Cox}, \citenamefont {d'~Aubign{\'{e}}}, \citenamefont
  {Bassani}, \citenamefont {Saminadayar},\ and\ \citenamefont
  {Tatarenko}}]{Kheng1993}%
  \BibitemOpen
  \bibfield  {author} {\bibinfo {author} {\bibfnamefont {K.}~\bibnamefont
  {Kheng}}, \bibinfo {author} {\bibfnamefont {R.}~\bibnamefont {Cox}}, \bibinfo
  {author} {\bibfnamefont {M.}~\bibnamefont {d'~Aubign{\'{e}}}}, \bibinfo
  {author} {\bibfnamefont {F.}~\bibnamefont {Bassani}}, \bibinfo {author}
  {\bibfnamefont {K.}~\bibnamefont {Saminadayar}},\ and\ \bibinfo {author}
  {\bibfnamefont {S.}~\bibnamefont {Tatarenko}},\ }\bibfield  {title} {\bibinfo
  {title} {{Observation of negatively charged excitons X- in semiconductor
  quantum wells}},\ }\href {https://doi.org/10.1103/PhysRevLett.71.1752}
  {\bibfield  {journal} {\bibinfo  {journal} {Phys. Rev. Lett.}\ }\textbf
  {\bibinfo {volume} {71}},\ \bibinfo {pages} {1752} (\bibinfo {year}
  {1993})}\BibitemShut {NoStop}%
\bibitem [{\citenamefont {Koudinov}\ \emph {et~al.}(2014)\citenamefont
  {Koudinov}, \citenamefont {Kehl}, \citenamefont {Rodina}, \citenamefont
  {Geurts}, \citenamefont {Wolverson},\ and\ \citenamefont
  {Karczewski}}]{Koudinov2014}%
  \BibitemOpen
  \bibfield  {author} {\bibinfo {author} {\bibfnamefont {A.~V.}\ \bibnamefont
  {Koudinov}}, \bibinfo {author} {\bibfnamefont {C.}~\bibnamefont {Kehl}},
  \bibinfo {author} {\bibfnamefont {A.~V.}\ \bibnamefont {Rodina}}, \bibinfo
  {author} {\bibfnamefont {J.}~\bibnamefont {Geurts}}, \bibinfo {author}
  {\bibfnamefont {D.}~\bibnamefont {Wolverson}},\ and\ \bibinfo {author}
  {\bibfnamefont {G.}~\bibnamefont {Karczewski}},\ }\bibfield  {title}
  {\bibinfo {title} {{Suris Tetrons: Possible Spectroscopic Evidence for
  Four-Particle Optical Excitations of a Two-Dimensional Electron Gas}},\
  }\href {https://doi.org/10.1103/PhysRevLett.112.147402} {\bibfield  {journal}
  {\bibinfo  {journal} {Phys. Rev. Lett.}\ }\textbf {\bibinfo {volume} {112}},\
  \bibinfo {pages} {147402} (\bibinfo {year} {2014})}\BibitemShut {NoStop}%
\bibitem [{\citenamefont {Sidler}\ \emph {et~al.}(2017)\citenamefont {Sidler},
  \citenamefont {Back}, \citenamefont {Cotlet}, \citenamefont {Srivastava},
  \citenamefont {Fink}, \citenamefont {Kroner}, \citenamefont {Demler},\ and\
  \citenamefont {Imamoglu}}]{Sidler2016}%
  \BibitemOpen
  \bibfield  {author} {\bibinfo {author} {\bibfnamefont {M.}~\bibnamefont
  {Sidler}}, \bibinfo {author} {\bibfnamefont {P.}~\bibnamefont {Back}},
  \bibinfo {author} {\bibfnamefont {O.}~\bibnamefont {Cotlet}}, \bibinfo
  {author} {\bibfnamefont {A.}~\bibnamefont {Srivastava}}, \bibinfo {author}
  {\bibfnamefont {T.}~\bibnamefont {Fink}}, \bibinfo {author} {\bibfnamefont
  {M.}~\bibnamefont {Kroner}}, \bibinfo {author} {\bibfnamefont
  {E.}~\bibnamefont {Demler}},\ and\ \bibinfo {author} {\bibfnamefont
  {A.}~\bibnamefont {Imamoglu}},\ }\bibfield  {title} {\bibinfo {title} {{Fermi
  polaron-polaritons in charge-tunable atomically thin semiconductors}},\
  }\href {https://doi.org/10.1038/nphys3949} {\bibfield  {journal} {\bibinfo
  {journal} {Nat. Phys.}\ }\textbf {\bibinfo {volume} {13}},\ \bibinfo {pages}
  {255} (\bibinfo {year} {2017})}\BibitemShut {NoStop}%
\bibitem [{\citenamefont {Efimkin}\ and\ \citenamefont
  {MacDonald}(2017)}]{Efimkin2017}%
  \BibitemOpen
  \bibfield  {author} {\bibinfo {author} {\bibfnamefont {D.~K.}\ \bibnamefont
  {Efimkin}}\ and\ \bibinfo {author} {\bibfnamefont {A.~H.}\ \bibnamefont
  {MacDonald}},\ }\bibfield  {title} {\bibinfo {title} {{Many-body theory of
  trion absorption features in two-dimensional semiconductors}},\ }\href
  {https://doi.org/10.1103/PhysRevB.95.035417} {\bibfield  {journal} {\bibinfo
  {journal} {Phys. Rev. B}\ }\textbf {\bibinfo {volume} {95}},\ \bibinfo
  {pages} {035417} (\bibinfo {year} {2017})},\ \Eprint
  {https://arxiv.org/abs/1609.06329} {arXiv:1609.06329} \BibitemShut {NoStop}%
\bibitem [{\citenamefont {Esser}\ \emph {et~al.}(2001)\citenamefont {Esser},
  \citenamefont {Zimmermann},\ and\ \citenamefont {Runge}}]{Esser2001}%
  \BibitemOpen
  \bibfield  {author} {\bibinfo {author} {\bibfnamefont {A.}~\bibnamefont
  {Esser}}, \bibinfo {author} {\bibfnamefont {R.}~\bibnamefont {Zimmermann}},\
  and\ \bibinfo {author} {\bibfnamefont {E.}~\bibnamefont {Runge}},\ }\bibfield
   {title} {\bibinfo {title} {{Theory of Trion Spectra in Semiconductor
  Nanostructures}},\ }\href
  {https://doi.org/10.1002/1521-3951(200110)227:2<317::AID-PSSB317>3.0.CO;2-S}
  {\bibfield  {journal} {\bibinfo  {journal} {Phys. status solidi}\ }\textbf
  {\bibinfo {volume} {227}},\ \bibinfo {pages} {317} (\bibinfo {year}
  {2001})}\BibitemShut {NoStop}%
\bibitem [{\citenamefont {Ivchenko}(2005)}]{Ivchenko2005}%
  \BibitemOpen
  \bibfield  {author} {\bibinfo {author} {\bibfnamefont {E.~L.}\ \bibnamefont
  {Ivchenko}},\ }\href@noop {} {\emph {\bibinfo {title} {{Optical spectroscopy
  of semiconductor nanostructures}}}}\ (\bibinfo  {publisher} {Alpha Science,
  Harrow England},\ \bibinfo {year} {2005})\BibitemShut {NoStop}%
\bibitem [{\citenamefont {Esser}\ \emph {et~al.}(2000)\citenamefont {Esser},
  \citenamefont {Runge}, \citenamefont {Zimmermann},\ and\ \citenamefont
  {Langbein}}]{Esser2000a}%
  \BibitemOpen
  \bibfield  {author} {\bibinfo {author} {\bibfnamefont {A.}~\bibnamefont
  {Esser}}, \bibinfo {author} {\bibfnamefont {E.}~\bibnamefont {Runge}},
  \bibinfo {author} {\bibfnamefont {R.}~\bibnamefont {Zimmermann}},\ and\
  \bibinfo {author} {\bibfnamefont {W.}~\bibnamefont {Langbein}},\ }\bibfield
  {title} {\bibinfo {title} {{Photoluminescence and radiative lifetime of
  trions in GaAs quantum wells}},\ }\href
  {https://doi.org/10.1103/PhysRevB.62.8232} {\bibfield  {journal} {\bibinfo
  {journal} {Phys. Rev. B}\ }\textbf {\bibinfo {volume} {62}},\ \bibinfo
  {pages} {8232} (\bibinfo {year} {2000})}\BibitemShut {NoStop}%
\bibitem [{\citenamefont {Ross}\ \emph {et~al.}(2013)\citenamefont {Ross},
  \citenamefont {Wu}, \citenamefont {Yu}, \citenamefont {Ghimire},
  \citenamefont {Jones}, \citenamefont {Aivazian}, \citenamefont {Yan},
  \citenamefont {Mandrus}, \citenamefont {Xiao}, \citenamefont {Yao},\ and\
  \citenamefont {Xu}}]{Ross2013}%
  \BibitemOpen
  \bibfield  {author} {\bibinfo {author} {\bibfnamefont {J.~S.}\ \bibnamefont
  {Ross}}, \bibinfo {author} {\bibfnamefont {S.}~\bibnamefont {Wu}}, \bibinfo
  {author} {\bibfnamefont {H.}~\bibnamefont {Yu}}, \bibinfo {author}
  {\bibfnamefont {N.~J.}\ \bibnamefont {Ghimire}}, \bibinfo {author}
  {\bibfnamefont {A.~M.}\ \bibnamefont {Jones}}, \bibinfo {author}
  {\bibfnamefont {G.}~\bibnamefont {Aivazian}}, \bibinfo {author}
  {\bibfnamefont {J.}~\bibnamefont {Yan}}, \bibinfo {author} {\bibfnamefont
  {D.~G.}\ \bibnamefont {Mandrus}}, \bibinfo {author} {\bibfnamefont
  {D.}~\bibnamefont {Xiao}}, \bibinfo {author} {\bibfnamefont {W.}~\bibnamefont
  {Yao}},\ and\ \bibinfo {author} {\bibfnamefont {X.}~\bibnamefont {Xu}},\
  }\bibfield  {title} {\bibinfo {title} {{Electrical control of neutral and
  charged excitons in a monolayer semiconductor.}},\ }\href
  {https://doi.org/10.1038/ncomms2498} {\bibfield  {journal} {\bibinfo
  {journal} {Nat. Commun.}\ }\textbf {\bibinfo {volume} {4}},\ \bibinfo {pages}
  {1474} (\bibinfo {year} {2013})}\BibitemShut {NoStop}%
\bibitem [{\citenamefont {Yu}\ \emph {et~al.}(2015)\citenamefont {Yu},
  \citenamefont {Cui}, \citenamefont {Xu},\ and\ \citenamefont {Yao}}]{Yu2015}%
  \BibitemOpen
  \bibfield  {author} {\bibinfo {author} {\bibfnamefont {H.}~\bibnamefont
  {Yu}}, \bibinfo {author} {\bibfnamefont {X.}~\bibnamefont {Cui}}, \bibinfo
  {author} {\bibfnamefont {X.}~\bibnamefont {Xu}},\ and\ \bibinfo {author}
  {\bibfnamefont {W.}~\bibnamefont {Yao}},\ }\bibfield  {title} {\bibinfo
  {title} {{Valley excitons in two-dimensional semiconductors}},\ }\href
  {https://doi.org/10.1093/nsr/nwu078} {\bibfield  {journal} {\bibinfo
  {journal} {Natl. Sci. Rev.}\ }\textbf {\bibinfo {volume} {2}},\ \bibinfo
  {pages} {57} (\bibinfo {year} {2015})}\BibitemShut {NoStop}%
\bibitem [{\citenamefont {Wang}\ \emph {et~al.}(2018)\citenamefont {Wang},
  \citenamefont {Chernikov}, \citenamefont {Glazov}, \citenamefont {Heinz},
  \citenamefont {Marie}, \citenamefont {Amand},\ and\ \citenamefont
  {Urbaszek}}]{Wang2018}%
  \BibitemOpen
  \bibfield  {author} {\bibinfo {author} {\bibfnamefont {G.}~\bibnamefont
  {Wang}}, \bibinfo {author} {\bibfnamefont {A.}~\bibnamefont {Chernikov}},
  \bibinfo {author} {\bibfnamefont {M.~M.}\ \bibnamefont {Glazov}}, \bibinfo
  {author} {\bibfnamefont {T.~F.}\ \bibnamefont {Heinz}}, \bibinfo {author}
  {\bibfnamefont {X.}~\bibnamefont {Marie}}, \bibinfo {author} {\bibfnamefont
  {T.}~\bibnamefont {Amand}},\ and\ \bibinfo {author} {\bibfnamefont
  {B.}~\bibnamefont {Urbaszek}},\ }\bibfield  {title} {\bibinfo {title}
  {{Colloquium : Excitons in atomically thin transition metal
  dichalcogenides}},\ }\href {https://doi.org/10.1103/RevModPhys.90.021001}
  {\bibfield  {journal} {\bibinfo  {journal} {Rev. Mod. Phys.}\ }\textbf
  {\bibinfo {volume} {90}},\ \bibinfo {pages} {021001} (\bibinfo {year}
  {2018})}\BibitemShut {NoStop}%
\bibitem [{\citenamefont {Christopher}\ \emph {et~al.}(2017)\citenamefont
  {Christopher}, \citenamefont {Goldberg},\ and\ \citenamefont
  {Swan}}]{Christopher2017}%
  \BibitemOpen
  \bibfield  {author} {\bibinfo {author} {\bibfnamefont {J.~W.}\ \bibnamefont
  {Christopher}}, \bibinfo {author} {\bibfnamefont {B.~B.}\ \bibnamefont
  {Goldberg}},\ and\ \bibinfo {author} {\bibfnamefont {A.~K.}\ \bibnamefont
  {Swan}},\ }\bibfield  {title} {\bibinfo {title} {{Long tailed trions in
  monolayer MoS2: Temperature dependent asymmetry and resulting red-shift of
  trion photoluminescence spectra}},\ }\href
  {https://doi.org/10.1038/s41598-017-14378-w} {\bibfield  {journal} {\bibinfo
  {journal} {Sci. Rep.}\ }\textbf {\bibinfo {volume} {7}},\ \bibinfo {pages}
  {14062} (\bibinfo {year} {2017})}\BibitemShut {NoStop}%
\bibitem [{\citenamefont {Lyons}\ \emph {et~al.}(2019)\citenamefont {Lyons},
  \citenamefont {Dufferwiel}, \citenamefont {Brooks}, \citenamefont {Withers},
  \citenamefont {Taniguchi}, \citenamefont {Watanabe}, \citenamefont
  {Novoselov}, \citenamefont {Burkard},\ and\ \citenamefont
  {Tartakovskii}}]{Lyons2019}%
  \BibitemOpen
  \bibfield  {author} {\bibinfo {author} {\bibfnamefont {T.~P.}\ \bibnamefont
  {Lyons}}, \bibinfo {author} {\bibfnamefont {S.}~\bibnamefont {Dufferwiel}},
  \bibinfo {author} {\bibfnamefont {M.}~\bibnamefont {Brooks}}, \bibinfo
  {author} {\bibfnamefont {F.}~\bibnamefont {Withers}}, \bibinfo {author}
  {\bibfnamefont {T.}~\bibnamefont {Taniguchi}}, \bibinfo {author}
  {\bibfnamefont {K.}~\bibnamefont {Watanabe}}, \bibinfo {author}
  {\bibfnamefont {K.~S.}\ \bibnamefont {Novoselov}}, \bibinfo {author}
  {\bibfnamefont {G.}~\bibnamefont {Burkard}},\ and\ \bibinfo {author}
  {\bibfnamefont {A.~I.}\ \bibnamefont {Tartakovskii}},\ }\bibfield  {title}
  {\bibinfo {title} {{The valley Zeeman effect in inter- and intra-valley
  trions in monolayer WSe2}},\ }\href
  {https://doi.org/10.1038/s41467-019-10228-7} {\bibfield  {journal} {\bibinfo
  {journal} {Nat. Commun.}\ }\textbf {\bibinfo {volume} {10}},\ \bibinfo
  {pages} {2330} (\bibinfo {year} {2019})}\BibitemShut {NoStop}%
\bibitem [{\citenamefont {Zhumagulov}\ \emph {et~al.}(2020)\citenamefont
  {Zhumagulov}, \citenamefont {Vagov}, \citenamefont {Gulevich}, \citenamefont
  {{Faria Junior}},\ and\ \citenamefont {Perebeinos}}]{Zhumagulov2020}%
  \BibitemOpen
  \bibfield  {author} {\bibinfo {author} {\bibfnamefont {Y.~V.}\ \bibnamefont
  {Zhumagulov}}, \bibinfo {author} {\bibfnamefont {A.}~\bibnamefont {Vagov}},
  \bibinfo {author} {\bibfnamefont {D.~R.}\ \bibnamefont {Gulevich}}, \bibinfo
  {author} {\bibfnamefont {P.~E.}\ \bibnamefont {{Faria Junior}}},\ and\
  \bibinfo {author} {\bibfnamefont {V.}~\bibnamefont {Perebeinos}},\ }\bibfield
   {title} {\bibinfo {title} {{Trion induced photoluminescence of a doped
  MoS2monolayer}},\ }\href {https://doi.org/10.1063/5.0012971} {\bibfield
  {journal} {\bibinfo  {journal} {J. Chem. Phys.}\ }\textbf {\bibinfo {volume}
  {153}},\ \bibinfo {pages} {1} (\bibinfo {year} {2020})},\ \Eprint
  {https://arxiv.org/abs/2005.09306} {arXiv:2005.09306} \BibitemShut {NoStop}%
\bibitem [{\citenamefont {Park}\ \emph {et~al.}(2021)\citenamefont {Park},
  \citenamefont {Han}, \citenamefont {Boule}, \citenamefont {Paget},
  \citenamefont {Rowe}, \citenamefont {Sirotti}, \citenamefont {Taniguchi},
  \citenamefont {Watanabe}, \citenamefont {Robert}, \citenamefont {Lombez},
  \citenamefont {Urbaszek}, \citenamefont {Marie},\ and\ \citenamefont
  {Cadiz}}]{Park2021}%
  \BibitemOpen
  \bibfield  {author} {\bibinfo {author} {\bibfnamefont {S.}~\bibnamefont
  {Park}}, \bibinfo {author} {\bibfnamefont {B.}~\bibnamefont {Han}}, \bibinfo
  {author} {\bibfnamefont {C.}~\bibnamefont {Boule}}, \bibinfo {author}
  {\bibfnamefont {D.}~\bibnamefont {Paget}}, \bibinfo {author} {\bibfnamefont
  {A.~C.~H.}\ \bibnamefont {Rowe}}, \bibinfo {author} {\bibfnamefont
  {F.}~\bibnamefont {Sirotti}}, \bibinfo {author} {\bibfnamefont
  {T.}~\bibnamefont {Taniguchi}}, \bibinfo {author} {\bibfnamefont
  {K.}~\bibnamefont {Watanabe}}, \bibinfo {author} {\bibfnamefont
  {C.}~\bibnamefont {Robert}}, \bibinfo {author} {\bibfnamefont
  {L.}~\bibnamefont {Lombez}}, \bibinfo {author} {\bibfnamefont
  {B.}~\bibnamefont {Urbaszek}}, \bibinfo {author} {\bibfnamefont
  {X.}~\bibnamefont {Marie}},\ and\ \bibinfo {author} {\bibfnamefont
  {F.}~\bibnamefont {Cadiz}},\ }\bibfield  {title} {\bibinfo {title} {{Imaging
  Seebeck drift of excitons and trions in MoSe 2 monolayers}},\ }\href
  {https://doi.org/10.1088/2053-1583/ac171f} {\bibfield  {journal} {\bibinfo
  {journal} {2D Mater.}\ }\textbf {\bibinfo {volume} {8}},\ \bibinfo {pages}
  {045014} (\bibinfo {year} {2021})}\BibitemShut {NoStop}%
\bibitem [{\citenamefont {Venanzi}\ \emph {et~al.}(2021)\citenamefont
  {Venanzi}, \citenamefont {Selig}, \citenamefont {Winnerl}, \citenamefont
  {Pashkin}, \citenamefont {Knorr}, \citenamefont {Helm},\ and\ \citenamefont
  {Schneider}}]{Venanzi2021}%
  \BibitemOpen
  \bibfield  {author} {\bibinfo {author} {\bibfnamefont {T.}~\bibnamefont
  {Venanzi}}, \bibinfo {author} {\bibfnamefont {M.}~\bibnamefont {Selig}},
  \bibinfo {author} {\bibfnamefont {S.}~\bibnamefont {Winnerl}}, \bibinfo
  {author} {\bibfnamefont {A.}~\bibnamefont {Pashkin}}, \bibinfo {author}
  {\bibfnamefont {A.}~\bibnamefont {Knorr}}, \bibinfo {author} {\bibfnamefont
  {M.}~\bibnamefont {Helm}},\ and\ \bibinfo {author} {\bibfnamefont
  {H.}~\bibnamefont {Schneider}},\ }\bibfield  {title} {\bibinfo {title}
  {{Terahertz-Induced Energy Transfer from Hot Carriers to Trions in a MoSe 2
  Monolayer}},\ }\href {https://doi.org/10.1021/acsphotonics.1c00394}
  {\bibfield  {journal} {\bibinfo  {journal} {ACS Photonics}\ ,\ \bibinfo
  {pages} {acsphotonics.1c00394}} (\bibinfo {year} {2021})}\BibitemShut
  {NoStop}%
\bibitem [{\citenamefont {Chang}\ \emph {et~al.}(2018)\citenamefont {Chang},
  \citenamefont {Shiau},\ and\ \citenamefont {Combescot}}]{Chang2018}%
  \BibitemOpen
  \bibfield  {author} {\bibinfo {author} {\bibfnamefont {Y.-C.}\ \bibnamefont
  {Chang}}, \bibinfo {author} {\bibfnamefont {S.-Y.}\ \bibnamefont {Shiau}},\
  and\ \bibinfo {author} {\bibfnamefont {M.}~\bibnamefont {Combescot}},\
  }\bibfield  {title} {\bibinfo {title} {{Crossover from trion-hole complex to
  exciton-polaron in n -doped two-dimensional semiconductor quantum wells}},\
  }\href {https://doi.org/10.1103/PhysRevB.98.235203} {\bibfield  {journal}
  {\bibinfo  {journal} {Phys. Rev. B}\ }\textbf {\bibinfo {volume} {98}},\
  \bibinfo {pages} {235203} (\bibinfo {year} {2018})}\BibitemShut {NoStop}%
\bibitem [{\citenamefont {Fey}\ \emph {et~al.}(2020)\citenamefont {Fey},
  \citenamefont {Schmelcher}, \citenamefont {Imamoglu},\ and\ \citenamefont
  {Schmidt}}]{Fey2020}%
  \BibitemOpen
  \bibfield  {author} {\bibinfo {author} {\bibfnamefont {C.}~\bibnamefont
  {Fey}}, \bibinfo {author} {\bibfnamefont {P.}~\bibnamefont {Schmelcher}},
  \bibinfo {author} {\bibfnamefont {A.}~\bibnamefont {Imamoglu}},\ and\
  \bibinfo {author} {\bibfnamefont {R.}~\bibnamefont {Schmidt}},\ }\bibfield
  {title} {\bibinfo {title} {{Theory of exciton-electron scattering in
  atomically thin semiconductors}},\ }\href
  {https://doi.org/10.1103/PhysRevB.101.195417} {\bibfield  {journal} {\bibinfo
   {journal} {Phys. Rev. B}\ }\textbf {\bibinfo {volume} {101}},\ \bibinfo
  {pages} {195417} (\bibinfo {year} {2020})}\BibitemShut {NoStop}%
\bibitem [{\citenamefont {Cotlet}\ \emph {et~al.}(2020)\citenamefont {Cotlet},
  \citenamefont {Wild}, \citenamefont {Lukin},\ and\ \citenamefont
  {Imamoglu}}]{Cotlet2020}%
  \BibitemOpen
  \bibfield  {author} {\bibinfo {author} {\bibfnamefont {O.}~\bibnamefont
  {Cotlet}}, \bibinfo {author} {\bibfnamefont {D.~S.}\ \bibnamefont {Wild}},
  \bibinfo {author} {\bibfnamefont {M.~D.}\ \bibnamefont {Lukin}},\ and\
  \bibinfo {author} {\bibfnamefont {A.}~\bibnamefont {Imamoglu}},\ }\bibfield
  {title} {\bibinfo {title} {{Rotons in optical excitation spectra of monolayer
  semiconductors}},\ }\href {https://doi.org/10.1103/PhysRevB.101.205409}
  {\bibfield  {journal} {\bibinfo  {journal} {Phys. Rev. B}\ }\textbf {\bibinfo
  {volume} {101}},\ \bibinfo {pages} {205409} (\bibinfo {year}
  {2020})}\BibitemShut {NoStop}%
\bibitem [{\citenamefont {Glazov}(2020{\natexlab{a}})}]{Glazov2020}%
  \BibitemOpen
  \bibfield  {author} {\bibinfo {author} {\bibfnamefont {M.~M.}\ \bibnamefont
  {Glazov}},\ }\bibfield  {title} {\bibinfo {title} {{Quantum Interference
  Effect on Exciton Transport in Monolayer Semiconductors}},\ }\href
  {https://doi.org/10.1103/PhysRevLett.124.166802} {\bibfield  {journal}
  {\bibinfo  {journal} {Phys. Rev. Lett.}\ }\textbf {\bibinfo {volume} {124}},\
  \bibinfo {pages} {166802} (\bibinfo {year} {2020}{\natexlab{a}})}\BibitemShut
  {NoStop}%
\bibitem [{Note1()}]{Note1}%
  \BibitemOpen
  \bibinfo {note} {Please also see a recent preprint discussing spectral
  lineshape asymmetry for Fermi polarons: T. Wasak et al.
  arXiv:2103.14040}\BibitemShut {NoStop}%
\bibitem [{\citenamefont {Castellanos-Gomez}\ \emph {et~al.}(2014)\citenamefont
  {Castellanos-Gomez}, \citenamefont {Vicarelli}, \citenamefont {Prada},
  \citenamefont {Island}, \citenamefont {Narasimha-Acharya}, \citenamefont
  {Blanter}, \citenamefont {Groenendijk}, \citenamefont {Buscema},
  \citenamefont {Steele}, \citenamefont {Alvarez}, \citenamefont {Zandbergen},
  \citenamefont {Palacios},\ and\ \citenamefont {van~der
  Zant}}]{Castellanos-Gomez2014}%
  \BibitemOpen
  \bibfield  {author} {\bibinfo {author} {\bibfnamefont {A.}~\bibnamefont
  {Castellanos-Gomez}}, \bibinfo {author} {\bibfnamefont {L.}~\bibnamefont
  {Vicarelli}}, \bibinfo {author} {\bibfnamefont {E.}~\bibnamefont {Prada}},
  \bibinfo {author} {\bibfnamefont {J.~O.}\ \bibnamefont {Island}}, \bibinfo
  {author} {\bibfnamefont {K.~L.}\ \bibnamefont {Narasimha-Acharya}}, \bibinfo
  {author} {\bibfnamefont {S.~I.}\ \bibnamefont {Blanter}}, \bibinfo {author}
  {\bibfnamefont {D.~J.}\ \bibnamefont {Groenendijk}}, \bibinfo {author}
  {\bibfnamefont {M.}~\bibnamefont {Buscema}}, \bibinfo {author} {\bibfnamefont
  {G.~a.}\ \bibnamefont {Steele}}, \bibinfo {author} {\bibfnamefont {J.~V.}\
  \bibnamefont {Alvarez}}, \bibinfo {author} {\bibfnamefont {H.~W.}\
  \bibnamefont {Zandbergen}}, \bibinfo {author} {\bibfnamefont {J.~J.}\
  \bibnamefont {Palacios}},\ and\ \bibinfo {author} {\bibfnamefont {H.~S.~J.}\
  \bibnamefont {van~der Zant}},\ }\bibfield  {title} {\bibinfo {title}
  {{Isolation and characterization of few-layer black phosphorus}},\ }\href
  {https://doi.org/10.1088/2053-1583/1/2/025001} {\bibfield  {journal}
  {\bibinfo  {journal} {2D Mater.}\ }\textbf {\bibinfo {volume} {1}},\ \bibinfo
  {pages} {025001} (\bibinfo {year} {2014})}\BibitemShut {NoStop}%
\bibitem [{\citenamefont {Wagner}\ \emph
  {et~al.}(2020{\natexlab{a}})\citenamefont {Wagner}, \citenamefont {Wietek},
  \citenamefont {Ziegler}, \citenamefont {Semina}, \citenamefont {Taniguchi},
  \citenamefont {Watanabe}, \citenamefont {Zipfel}, \citenamefont {Glazov},\
  and\ \citenamefont {Chernikov}}]{Wagner2020}%
  \BibitemOpen
  \bibfield  {author} {\bibinfo {author} {\bibfnamefont {K.}~\bibnamefont
  {Wagner}}, \bibinfo {author} {\bibfnamefont {E.}~\bibnamefont {Wietek}},
  \bibinfo {author} {\bibfnamefont {J.~D.}\ \bibnamefont {Ziegler}}, \bibinfo
  {author} {\bibfnamefont {M.~A.}\ \bibnamefont {Semina}}, \bibinfo {author}
  {\bibfnamefont {T.}~\bibnamefont {Taniguchi}}, \bibinfo {author}
  {\bibfnamefont {K.}~\bibnamefont {Watanabe}}, \bibinfo {author}
  {\bibfnamefont {J.}~\bibnamefont {Zipfel}}, \bibinfo {author} {\bibfnamefont
  {M.~M.}\ \bibnamefont {Glazov}},\ and\ \bibinfo {author} {\bibfnamefont
  {A.}~\bibnamefont {Chernikov}},\ }\bibfield  {title} {\bibinfo {title}
  {{Autoionization and Dressing of Excited Excitons by Free Carriers in
  Monolayer WSe2}},\ }\href {https://doi.org/10.1103/PhysRevLett.125.267401}
  {\bibfield  {journal} {\bibinfo  {journal} {Phys. Rev. Lett.}\ }\textbf
  {\bibinfo {volume} {125}},\ \bibinfo {pages} {267401} (\bibinfo {year}
  {2020}{\natexlab{a}})}\BibitemShut {NoStop}%
\bibitem [{\citenamefont {Smole{\'{n}}ski}\ \emph {et~al.}(2019)\citenamefont
  {Smole{\'{n}}ski}, \citenamefont {Cotlet}, \citenamefont {Popert},
  \citenamefont {Back}, \citenamefont {Shimazaki}, \citenamefont
  {Kn{\"{u}}ppel}, \citenamefont {Dietler}, \citenamefont {Taniguchi},
  \citenamefont {Watanabe}, \citenamefont {Kroner},\ and\ \citenamefont
  {Imamoglu}}]{Smolenski2019}%
  \BibitemOpen
  \bibfield  {author} {\bibinfo {author} {\bibfnamefont {T.}~\bibnamefont
  {Smole{\'{n}}ski}}, \bibinfo {author} {\bibfnamefont {O.}~\bibnamefont
  {Cotlet}}, \bibinfo {author} {\bibfnamefont {A.}~\bibnamefont {Popert}},
  \bibinfo {author} {\bibfnamefont {P.}~\bibnamefont {Back}}, \bibinfo {author}
  {\bibfnamefont {Y.}~\bibnamefont {Shimazaki}}, \bibinfo {author}
  {\bibfnamefont {P.}~\bibnamefont {Kn{\"{u}}ppel}}, \bibinfo {author}
  {\bibfnamefont {N.}~\bibnamefont {Dietler}}, \bibinfo {author} {\bibfnamefont
  {T.}~\bibnamefont {Taniguchi}}, \bibinfo {author} {\bibfnamefont
  {K.}~\bibnamefont {Watanabe}}, \bibinfo {author} {\bibfnamefont
  {M.}~\bibnamefont {Kroner}},\ and\ \bibinfo {author} {\bibfnamefont
  {A.}~\bibnamefont {Imamoglu}},\ }\bibfield  {title} {\bibinfo {title}
  {{Interaction-Induced Shubnikov--de Haas Oscillations in Optical Conductivity
  of Monolayer MoSe2}},\ }\href
  {https://doi.org/10.1103/PhysRevLett.123.097403} {\bibfield  {journal}
  {\bibinfo  {journal} {Phys. Rev. Lett.}\ }\textbf {\bibinfo {volume} {123}},\
  \bibinfo {pages} {097403} (\bibinfo {year} {2019})}\BibitemShut {NoStop}%
\bibitem [{\citenamefont {Shepard}\ \emph {et~al.}(2017)\citenamefont
  {Shepard}, \citenamefont {Ardelean}, \citenamefont {Ajayi}, \citenamefont
  {Rhodes}, \citenamefont {Zhu}, \citenamefont {Hone},\ and\ \citenamefont
  {Strauf}}]{Shepard2017}%
  \BibitemOpen
  \bibfield  {author} {\bibinfo {author} {\bibfnamefont {G.~D.}\ \bibnamefont
  {Shepard}}, \bibinfo {author} {\bibfnamefont {J.~V.}\ \bibnamefont
  {Ardelean}}, \bibinfo {author} {\bibfnamefont {O.~A.}\ \bibnamefont {Ajayi}},
  \bibinfo {author} {\bibfnamefont {D.}~\bibnamefont {Rhodes}}, \bibinfo
  {author} {\bibfnamefont {X.}~\bibnamefont {Zhu}}, \bibinfo {author}
  {\bibfnamefont {J.~C.}\ \bibnamefont {Hone}},\ and\ \bibinfo {author}
  {\bibfnamefont {S.}~\bibnamefont {Strauf}},\ }\bibfield  {title} {\bibinfo
  {title} {{Trion-Species-Resolved Quantum Beats in MoSe 2}},\ }\href
  {https://doi.org/10.1021/acsnano.7b06444} {\bibfield  {journal} {\bibinfo
  {journal} {ACS Nano}\ }\textbf {\bibinfo {volume} {11}},\ \bibinfo {pages}
  {11550} (\bibinfo {year} {2017})}\BibitemShut {NoStop}%
\bibitem [{\citenamefont {Rau}(1996)}]{Rau1996}%
  \BibitemOpen
  \bibfield  {author} {\bibinfo {author} {\bibfnamefont {A.~R.~P.}\
  \bibnamefont {Rau}},\ }\bibfield  {title} {\bibinfo {title} {{The Negative
  Ion of Hydrogen}},\ }\href@noop {} {\bibfield  {journal} {\bibinfo  {journal}
  {J. Astrophys. Astr.}\ }\textbf {\bibinfo {volume} {17}},\ \bibinfo {pages}
  {113} (\bibinfo {year} {1996})}\BibitemShut {NoStop}%
\bibitem [{\citenamefont {Ganchev}\ \emph {et~al.}(2015)\citenamefont
  {Ganchev}, \citenamefont {Drummond}, \citenamefont {Aleiner},\ and\
  \citenamefont {Fal'ko}}]{Ganchev2015}%
  \BibitemOpen
  \bibfield  {author} {\bibinfo {author} {\bibfnamefont {B.}~\bibnamefont
  {Ganchev}}, \bibinfo {author} {\bibfnamefont {N.}~\bibnamefont {Drummond}},
  \bibinfo {author} {\bibfnamefont {I.}~\bibnamefont {Aleiner}},\ and\ \bibinfo
  {author} {\bibfnamefont {V.}~\bibnamefont {Fal'ko}},\ }\bibfield  {title}
  {\bibinfo {title} {{Three-Particle Complexes in Two-Dimensional
  Semiconductors}},\ }\href {https://doi.org/10.1103/PhysRevLett.114.107401}
  {\bibfield  {journal} {\bibinfo  {journal} {Phys. Rev. Lett.}\ }\textbf
  {\bibinfo {volume} {114}},\ \bibinfo {pages} {107401} (\bibinfo {year}
  {2015})}\BibitemShut {NoStop}%
\bibitem [{\citenamefont {Glazov}(2020{\natexlab{b}})}]{Glazov2020c}%
  \BibitemOpen
  \bibfield  {author} {\bibinfo {author} {\bibfnamefont {M.~M.}\ \bibnamefont
  {Glazov}},\ }\bibfield  {title} {\bibinfo {title} {{Optical properties of
  charged excitons in two-dimensional semiconductors}},\ }\href
  {https://doi.org/10.1063/5.0012475} {\bibfield  {journal} {\bibinfo
  {journal} {J. Chem. Phys.}\ }\textbf {\bibinfo {volume} {153}},\ \bibinfo
  {pages} {034703} (\bibinfo {year} {2020}{\natexlab{b}})},\ \Eprint
  {https://arxiv.org/abs/2005.05829} {arXiv:2005.05829} \BibitemShut {NoStop}%
\bibitem [{\citenamefont {Suris}\ \emph {et~al.}(2001)\citenamefont {Suris},
  \citenamefont {Kochereshko}, \citenamefont {Astakhov}, \citenamefont
  {Yakovlev}, \citenamefont {Ossau}, \citenamefont {Nurnberger}, \citenamefont
  {Faschinger}, \citenamefont {Landwehr}, \citenamefont {Wojtowicz},
  \citenamefont {Karczewski},\ and\ \citenamefont {Kossut}}]{Suris2001}%
  \BibitemOpen
  \bibfield  {author} {\bibinfo {author} {\bibfnamefont {R.}~\bibnamefont
  {Suris}}, \bibinfo {author} {\bibfnamefont {V.}~\bibnamefont {Kochereshko}},
  \bibinfo {author} {\bibfnamefont {G.}~\bibnamefont {Astakhov}}, \bibinfo
  {author} {\bibfnamefont {D.}~\bibnamefont {Yakovlev}}, \bibinfo {author}
  {\bibfnamefont {W.}~\bibnamefont {Ossau}}, \bibinfo {author} {\bibfnamefont
  {J.}~\bibnamefont {Nurnberger}}, \bibinfo {author} {\bibfnamefont
  {W.}~\bibnamefont {Faschinger}}, \bibinfo {author} {\bibfnamefont
  {G.}~\bibnamefont {Landwehr}}, \bibinfo {author} {\bibfnamefont
  {T.}~\bibnamefont {Wojtowicz}}, \bibinfo {author} {\bibfnamefont
  {G.}~\bibnamefont {Karczewski}},\ and\ \bibinfo {author} {\bibfnamefont
  {J.}~\bibnamefont {Kossut}},\ }\bibfield  {title} {\bibinfo {title} {Excitons
  and trions modified by interaction with a two-dimensional electron gas},\
  }\href
  {https://doi.org/10.1002/1521-3951(200110)227:2<343::AID-PSSB343>3.0.CO;2-W}
  {\bibfield  {journal} {\bibinfo  {journal} {Phys. status solidi}\ }\textbf
  {\bibinfo {volume} {227}},\ \bibinfo {pages} {343} (\bibinfo {year}
  {2001})}\BibitemShut {NoStop}%
\bibitem [{\citenamefont {Suris}(2003)}]{suris:correlation}%
  \BibitemOpen
  \bibfield  {author} {\bibinfo {author} {\bibfnamefont {R.~A.}\ \bibnamefont
  {Suris}},\ }\bibinfo {title} {Optical properties of 2d systems with
  interacting electrons}\ (\bibinfo  {publisher} {NATO ASI},\ \bibinfo {year}
  {2003})\ Chap.\ \bibinfo {chapter} {Correlation between trion and hole in
  fermi distribution in process of trion photo-excitation in doped
  QWs}\BibitemShut {NoStop}%
\bibitem [{\citenamefont {Schmidt}\ \emph {et~al.}(2012)\citenamefont
  {Schmidt}, \citenamefont {Enss}, \citenamefont {Pietil\"a},\ and\
  \citenamefont {Demler}}]{PhysRevA.85.021602}%
  \BibitemOpen
  \bibfield  {author} {\bibinfo {author} {\bibfnamefont {R.}~\bibnamefont
  {Schmidt}}, \bibinfo {author} {\bibfnamefont {T.}~\bibnamefont {Enss}},
  \bibinfo {author} {\bibfnamefont {V.}~\bibnamefont {Pietil\"a}},\ and\
  \bibinfo {author} {\bibfnamefont {E.}~\bibnamefont {Demler}},\ }\bibfield
  {title} {\bibinfo {title} {Fermi polarons in two dimensions},\ }\href
  {https://doi.org/10.1103/PhysRevA.85.021602} {\bibfield  {journal} {\bibinfo
  {journal} {Phys. Rev. A}\ }\textbf {\bibinfo {volume} {85}},\ \bibinfo
  {pages} {021602} (\bibinfo {year} {2012})}\BibitemShut {NoStop}%
\bibitem [{\citenamefont {Cotlet}\ \emph {et~al.}(2019)\citenamefont {Cotlet},
  \citenamefont {Pientka}, \citenamefont {Schmidt}, \citenamefont {Zarand},
  \citenamefont {Demler},\ and\ \citenamefont {Imamoglu}}]{Cotlet2019}%
  \BibitemOpen
  \bibfield  {author} {\bibinfo {author} {\bibfnamefont {O.}~\bibnamefont
  {Cotlet}}, \bibinfo {author} {\bibfnamefont {F.}~\bibnamefont {Pientka}},
  \bibinfo {author} {\bibfnamefont {R.}~\bibnamefont {Schmidt}}, \bibinfo
  {author} {\bibfnamefont {G.}~\bibnamefont {Zarand}}, \bibinfo {author}
  {\bibfnamefont {E.}~\bibnamefont {Demler}},\ and\ \bibinfo {author}
  {\bibfnamefont {A.}~\bibnamefont {Imamoglu}},\ }\bibfield  {title} {\bibinfo
  {title} {{Transport of Neutral Optical Excitations Using Electric Fields}},\
  }\href {https://doi.org/10.1103/PhysRevX.9.041019} {\bibfield  {journal}
  {\bibinfo  {journal} {Phys. Rev. X}\ }\textbf {\bibinfo {volume} {9}},\
  \bibinfo {pages} {041019} (\bibinfo {year} {2019})}\BibitemShut {NoStop}%
\bibitem [{Note2()}]{Note2}%
  \BibitemOpen
  \bibinfo {note} {Here we correct a typo in Eqs. (24), (26) of Ref. \cite
  {Glazov2020c} where the factor $m_x/m_{tr}$ was missing in the
  exponent.}\BibitemShut {Stop}%
\bibitem [{\citenamefont {Sergeev}\ and\ \citenamefont
  {Suris}(2001)}]{Sergeev2001}%
  \BibitemOpen
  \bibfield  {author} {\bibinfo {author} {\bibfnamefont {R.~A.}\ \bibnamefont
  {Sergeev}}\ and\ \bibinfo {author} {\bibfnamefont {R.~A.}\ \bibnamefont
  {Suris}},\ }\bibfield  {title} {\bibinfo {title} {Ground-state energy of
  $x^-$ and $x^+$ trions in a two-dimensional quantum well at an arbitrary mass
  ratio},\ }\href {https://doi.org/10.1134/1.1366005} {\bibfield  {journal}
  {\bibinfo  {journal} {Phys. Solid State}\ }\textbf {\bibinfo {volume} {43}},\
  \bibinfo {pages} {746} (\bibinfo {year} {2001})}\BibitemShut {NoStop}%
\bibitem [{\citenamefont {Berkelbach}\ \emph {et~al.}(2013)\citenamefont
  {Berkelbach}, \citenamefont {Hybertsen},\ and\ \citenamefont
  {Reichman}}]{Berkelbach2013}%
  \BibitemOpen
  \bibfield  {author} {\bibinfo {author} {\bibfnamefont {T.~C.}\ \bibnamefont
  {Berkelbach}}, \bibinfo {author} {\bibfnamefont {M.~S.}\ \bibnamefont
  {Hybertsen}},\ and\ \bibinfo {author} {\bibfnamefont {D.~R.}\ \bibnamefont
  {Reichman}},\ }\bibfield  {title} {\bibinfo {title} {{Theory of neutral and
  charged excitons in monolayer transition metal dichalcogenides}},\ }\href
  {https://doi.org/10.1103/PhysRevB.88.045318} {\bibfield  {journal} {\bibinfo
  {journal} {Phys. Rev. B}\ }\textbf {\bibinfo {volume} {88}},\ \bibinfo
  {pages} {045318} (\bibinfo {year} {2013})}\BibitemShut {NoStop}%
\bibitem [{\citenamefont {Courtade}\ \emph {et~al.}(2017)\citenamefont
  {Courtade}, \citenamefont {Semina}, \citenamefont {Manca}, \citenamefont
  {Glazov}, \citenamefont {Robert}, \citenamefont {Cadiz}, \citenamefont
  {Wang}, \citenamefont {Taniguchi}, \citenamefont {Watanabe}, \citenamefont
  {Pierre}, \citenamefont {Escoffier}, \citenamefont {Ivchenko}, \citenamefont
  {Renucci}, \citenamefont {Marie}, \citenamefont {Amand},\ and\ \citenamefont
  {Urbaszek}}]{Courtade2017}%
  \BibitemOpen
  \bibfield  {author} {\bibinfo {author} {\bibfnamefont {E.}~\bibnamefont
  {Courtade}}, \bibinfo {author} {\bibfnamefont {M.}~\bibnamefont {Semina}},
  \bibinfo {author} {\bibfnamefont {M.}~\bibnamefont {Manca}}, \bibinfo
  {author} {\bibfnamefont {M.~M.}\ \bibnamefont {Glazov}}, \bibinfo {author}
  {\bibfnamefont {C.}~\bibnamefont {Robert}}, \bibinfo {author} {\bibfnamefont
  {F.}~\bibnamefont {Cadiz}}, \bibinfo {author} {\bibfnamefont
  {G.}~\bibnamefont {Wang}}, \bibinfo {author} {\bibfnamefont {T.}~\bibnamefont
  {Taniguchi}}, \bibinfo {author} {\bibfnamefont {K.}~\bibnamefont {Watanabe}},
  \bibinfo {author} {\bibfnamefont {M.}~\bibnamefont {Pierre}}, \bibinfo
  {author} {\bibfnamefont {W.}~\bibnamefont {Escoffier}}, \bibinfo {author}
  {\bibfnamefont {E.~L.}\ \bibnamefont {Ivchenko}}, \bibinfo {author}
  {\bibfnamefont {P.}~\bibnamefont {Renucci}}, \bibinfo {author} {\bibfnamefont
  {X.}~\bibnamefont {Marie}}, \bibinfo {author} {\bibfnamefont
  {T.}~\bibnamefont {Amand}},\ and\ \bibinfo {author} {\bibfnamefont
  {B.}~\bibnamefont {Urbaszek}},\ }\bibfield  {title} {\bibinfo {title}
  {{Charged excitons in monolayer WSe2 : Experiment and theory}},\ }\href
  {https://doi.org/10.1103/PhysRevB.96.085302} {\bibfield  {journal} {\bibinfo
  {journal} {Phys. Rev. B}\ }\textbf {\bibinfo {volume} {96}},\ \bibinfo
  {pages} {085302} (\bibinfo {year} {2017})},\ \Eprint
  {https://arxiv.org/abs/1705.02110} {arXiv:1705.02110} \BibitemShut {NoStop}%
\bibitem [{\citenamefont {Imamoglu}\ \emph {et~al.}(2020)\citenamefont
  {Imamoglu}, \citenamefont {Cotlet},\ and\ \citenamefont
  {Schmidt}}]{imamoglu2020excitonpolarons}%
  \BibitemOpen
  \bibfield  {author} {\bibinfo {author} {\bibfnamefont {A.}~\bibnamefont
  {Imamoglu}}, \bibinfo {author} {\bibfnamefont {O.}~\bibnamefont {Cotlet}},\
  and\ \bibinfo {author} {\bibfnamefont {R.}~\bibnamefont {Schmidt}},\
  }\href@noop {} {\bibinfo {title} {Exciton-polarons in two-dimensional
  semiconductors and the tavis-cummings model}} (\bibinfo {year} {2020}),\
  \Eprint {https://arxiv.org/abs/2006.15963} {arXiv:2006.15963
  [cond-mat.mes-hall]} \BibitemShut {NoStop}%
\bibitem [{\citenamefont {Christiansen}\ \emph {et~al.}(2017)\citenamefont
  {Christiansen}, \citenamefont {Selig}, \citenamefont {Bergh{\"{a}}user},
  \citenamefont {Schmidt}, \citenamefont {Niehues}, \citenamefont {Schneider},
  \citenamefont {Arora}, \citenamefont {de~Vasconcellos}, \citenamefont
  {Bratschitsch}, \citenamefont {Malic},\ and\ \citenamefont
  {Knorr}}]{Christiansen2017}%
  \BibitemOpen
  \bibfield  {author} {\bibinfo {author} {\bibfnamefont {D.}~\bibnamefont
  {Christiansen}}, \bibinfo {author} {\bibfnamefont {M.}~\bibnamefont {Selig}},
  \bibinfo {author} {\bibfnamefont {G.}~\bibnamefont {Bergh{\"{a}}user}},
  \bibinfo {author} {\bibfnamefont {R.}~\bibnamefont {Schmidt}}, \bibinfo
  {author} {\bibfnamefont {I.}~\bibnamefont {Niehues}}, \bibinfo {author}
  {\bibfnamefont {R.}~\bibnamefont {Schneider}}, \bibinfo {author}
  {\bibfnamefont {A.}~\bibnamefont {Arora}}, \bibinfo {author} {\bibfnamefont
  {S.~M.}\ \bibnamefont {de~Vasconcellos}}, \bibinfo {author} {\bibfnamefont
  {R.}~\bibnamefont {Bratschitsch}}, \bibinfo {author} {\bibfnamefont
  {E.}~\bibnamefont {Malic}},\ and\ \bibinfo {author} {\bibfnamefont
  {A.}~\bibnamefont {Knorr}},\ }\bibfield  {title} {\bibinfo {title} {{Phonon
  Sidebands in Monolayer Transition Metal Dichalcogenides}},\ }\href
  {https://doi.org/10.1103/PhysRevLett.119.187402} {\bibfield  {journal}
  {\bibinfo  {journal} {Phys. Rev. Lett.}\ }\textbf {\bibinfo {volume} {119}},\
  \bibinfo {pages} {187402} (\bibinfo {year} {2017})}\BibitemShut {NoStop}%
\bibitem [{\citenamefont {Shree}\ \emph {et~al.}(2018)\citenamefont {Shree},
  \citenamefont {Semina}, \citenamefont {Robert}, \citenamefont {Han},
  \citenamefont {Amand}, \citenamefont {Balocchi}, \citenamefont {Manca},
  \citenamefont {Courtade}, \citenamefont {Marie}, \citenamefont {Taniguchi},
  \citenamefont {Watanabe}, \citenamefont {Glazov},\ and\ \citenamefont
  {Urbaszek}}]{shree2018exciton}%
  \BibitemOpen
  \bibfield  {author} {\bibinfo {author} {\bibfnamefont {S.}~\bibnamefont
  {Shree}}, \bibinfo {author} {\bibfnamefont {M.}~\bibnamefont {Semina}},
  \bibinfo {author} {\bibfnamefont {C.}~\bibnamefont {Robert}}, \bibinfo
  {author} {\bibfnamefont {B.}~\bibnamefont {Han}}, \bibinfo {author}
  {\bibfnamefont {T.}~\bibnamefont {Amand}}, \bibinfo {author} {\bibfnamefont
  {A.}~\bibnamefont {Balocchi}}, \bibinfo {author} {\bibfnamefont
  {M.}~\bibnamefont {Manca}}, \bibinfo {author} {\bibfnamefont
  {E.}~\bibnamefont {Courtade}}, \bibinfo {author} {\bibfnamefont
  {X.}~\bibnamefont {Marie}}, \bibinfo {author} {\bibfnamefont
  {T.}~\bibnamefont {Taniguchi}}, \bibinfo {author} {\bibfnamefont
  {K.}~\bibnamefont {Watanabe}}, \bibinfo {author} {\bibfnamefont {M.~M.}\
  \bibnamefont {Glazov}},\ and\ \bibinfo {author} {\bibfnamefont
  {B.}~\bibnamefont {Urbaszek}},\ }\bibfield  {title} {\bibinfo {title}
  {Observation of exciton-phonon coupling in {MoSe}$_{2}$ monolayers},\ }\href
  {https://doi.org/10.1103/PhysRevB.98.035302} {\bibfield  {journal} {\bibinfo
  {journal} {Phys. Rev. B}\ }\textbf {\bibinfo {volume} {98}},\ \bibinfo
  {pages} {035302} (\bibinfo {year} {2018})}\BibitemShut {NoStop}%
\bibitem [{\citenamefont {Gantmakher}\ and\ \citenamefont
  {Levinson}(1987)}]{gantmakher87}%
  \BibitemOpen
  \bibfield  {author} {\bibinfo {author} {\bibfnamefont {V.~F.}\ \bibnamefont
  {Gantmakher}}\ and\ \bibinfo {author} {\bibfnamefont {Y.~B.}\ \bibnamefont
  {Levinson}},\ }\href@noop {} {\emph {\bibinfo {title} {Carrier Scattering in
  Metals and Semiconductors}}}\ (\bibinfo  {publisher} {North-Holland
  Publishing Company},\ \bibinfo {year} {1987})\BibitemShut {NoStop}%
\bibitem [{\citenamefont {Kaasbjerg}\ \emph {et~al.}(2012)\citenamefont
  {Kaasbjerg}, \citenamefont {Thygesen},\ and\ \citenamefont
  {Jacobsen}}]{Kaasbjerg2012}%
  \BibitemOpen
  \bibfield  {author} {\bibinfo {author} {\bibfnamefont {K.}~\bibnamefont
  {Kaasbjerg}}, \bibinfo {author} {\bibfnamefont {K.~S.}\ \bibnamefont
  {Thygesen}},\ and\ \bibinfo {author} {\bibfnamefont {K.~W.}\ \bibnamefont
  {Jacobsen}},\ }\bibfield  {title} {\bibinfo {title} {{Phonon-limited mobility
  in n-type single-layer MoS2 from first principles}},\ }\href
  {https://doi.org/10.1103/PhysRevB.85.115317} {\bibfield  {journal} {\bibinfo
  {journal} {Phys. Rev. B}\ }\textbf {\bibinfo {volume} {85}},\ \bibinfo
  {pages} {115317} (\bibinfo {year} {2012})}\BibitemShut {NoStop}%
\bibitem [{\citenamefont {Steinhoff}\ \emph {et~al.}(2014)\citenamefont
  {Steinhoff}, \citenamefont {R{\"{o}}sner}, \citenamefont {Jahnke},
  \citenamefont {Wehling},\ and\ \citenamefont {Gies}}]{Steinhoff2014}%
  \BibitemOpen
  \bibfield  {author} {\bibinfo {author} {\bibfnamefont {A.}~\bibnamefont
  {Steinhoff}}, \bibinfo {author} {\bibfnamefont {M.}~\bibnamefont
  {R{\"{o}}sner}}, \bibinfo {author} {\bibfnamefont {F.}~\bibnamefont
  {Jahnke}}, \bibinfo {author} {\bibfnamefont {T.~O.}\ \bibnamefont
  {Wehling}},\ and\ \bibinfo {author} {\bibfnamefont {C.}~\bibnamefont
  {Gies}},\ }\bibfield  {title} {\bibinfo {title} {{Influence of excited
  carriers on the optical and electronic properties of MoS$_2$.}},\ }\href
  {https://doi.org/10.1021/nl500595u} {\bibfield  {journal} {\bibinfo
  {journal} {Nano Lett.}\ }\textbf {\bibinfo {volume} {14}},\ \bibinfo {pages}
  {3743} (\bibinfo {year} {2014})}\BibitemShut {NoStop}%
\bibitem [{\citenamefont {Schmidt}\ \emph {et~al.}(2016)\citenamefont
  {Schmidt}, \citenamefont {Bergh{\"{a}}user}, \citenamefont {Schneider},
  \citenamefont {Selig}, \citenamefont {Tonndorf}, \citenamefont {Mali{\'{c}}},
  \citenamefont {Knorr}, \citenamefont {{Michaelis de Vasconcellos}},\ and\
  \citenamefont {Bratschitsch}}]{Schmidt2016}%
  \BibitemOpen
  \bibfield  {author} {\bibinfo {author} {\bibfnamefont {R.}~\bibnamefont
  {Schmidt}}, \bibinfo {author} {\bibfnamefont {G.}~\bibnamefont
  {Bergh{\"{a}}user}}, \bibinfo {author} {\bibfnamefont {R.}~\bibnamefont
  {Schneider}}, \bibinfo {author} {\bibfnamefont {M.}~\bibnamefont {Selig}},
  \bibinfo {author} {\bibfnamefont {P.}~\bibnamefont {Tonndorf}}, \bibinfo
  {author} {\bibfnamefont {E.}~\bibnamefont {Mali{\'{c}}}}, \bibinfo {author}
  {\bibfnamefont {A.}~\bibnamefont {Knorr}}, \bibinfo {author} {\bibfnamefont
  {S.}~\bibnamefont {{Michaelis de Vasconcellos}}},\ and\ \bibinfo {author}
  {\bibfnamefont {R.}~\bibnamefont {Bratschitsch}},\ }\bibfield  {title}
  {\bibinfo {title} {{Ultrafast Coulomb-Induced Intervalley Coupling in
  Atomically Thin WS 2}},\ }\href
  {https://doi.org/10.1021/acs.nanolett.5b04733} {\bibfield  {journal}
  {\bibinfo  {journal} {Nano Lett.}\ }\textbf {\bibinfo {volume} {16}},\
  \bibinfo {pages} {2945} (\bibinfo {year} {2016})}\BibitemShut {NoStop}%
\bibitem [{\citenamefont {Fang}\ \emph {et~al.}(2019)\citenamefont {Fang},
  \citenamefont {Han}, \citenamefont {Robert}, \citenamefont {Semina},
  \citenamefont {Lagarde}, \citenamefont {Courtade}, \citenamefont {Taniguchi},
  \citenamefont {Watanabe}, \citenamefont {Amand}, \citenamefont {Urbaszek},
  \citenamefont {Glazov},\ and\ \citenamefont {Marie}}]{Fang2019}%
  \BibitemOpen
  \bibfield  {author} {\bibinfo {author} {\bibfnamefont {H.~H.}\ \bibnamefont
  {Fang}}, \bibinfo {author} {\bibfnamefont {B.}~\bibnamefont {Han}}, \bibinfo
  {author} {\bibfnamefont {C.}~\bibnamefont {Robert}}, \bibinfo {author}
  {\bibfnamefont {M.~A.}\ \bibnamefont {Semina}}, \bibinfo {author}
  {\bibfnamefont {D.}~\bibnamefont {Lagarde}}, \bibinfo {author} {\bibfnamefont
  {E.}~\bibnamefont {Courtade}}, \bibinfo {author} {\bibfnamefont
  {T.}~\bibnamefont {Taniguchi}}, \bibinfo {author} {\bibfnamefont
  {K.}~\bibnamefont {Watanabe}}, \bibinfo {author} {\bibfnamefont
  {T.}~\bibnamefont {Amand}}, \bibinfo {author} {\bibfnamefont
  {B.}~\bibnamefont {Urbaszek}}, \bibinfo {author} {\bibfnamefont {M.~M.}\
  \bibnamefont {Glazov}},\ and\ \bibinfo {author} {\bibfnamefont
  {X.}~\bibnamefont {Marie}},\ }\bibfield  {title} {\bibinfo {title} {{Control
  of the Exciton Radiative Lifetime in van der Waals Heterostructures}},\
  }\href {https://doi.org/10.1103/PhysRevLett.123.067401} {\bibfield  {journal}
  {\bibinfo  {journal} {Phys. Rev. Lett.}\ }\textbf {\bibinfo {volume} {123}},\
  \bibinfo {pages} {067401} (\bibinfo {year} {2019})}\BibitemShut {NoStop}%
\bibitem [{\citenamefont {Rosati}\ \emph {et~al.}(2020)\citenamefont {Rosati},
  \citenamefont {Wagner}, \citenamefont {Brem}, \citenamefont
  {Perea-Caus{\'{i}}n}, \citenamefont {Wietek}, \citenamefont {Zipfel},
  \citenamefont {Ziegler}, \citenamefont {Selig}, \citenamefont {Taniguchi},
  \citenamefont {Watanabe}, \citenamefont {Knorr}, \citenamefont {Chernikov},\
  and\ \citenamefont {Malic}}]{Rosati2020}%
  \BibitemOpen
  \bibfield  {author} {\bibinfo {author} {\bibfnamefont {R.}~\bibnamefont
  {Rosati}}, \bibinfo {author} {\bibfnamefont {K.}~\bibnamefont {Wagner}},
  \bibinfo {author} {\bibfnamefont {S.}~\bibnamefont {Brem}}, \bibinfo {author}
  {\bibfnamefont {R.}~\bibnamefont {Perea-Caus{\'{i}}n}}, \bibinfo {author}
  {\bibfnamefont {E.}~\bibnamefont {Wietek}}, \bibinfo {author} {\bibfnamefont
  {J.}~\bibnamefont {Zipfel}}, \bibinfo {author} {\bibfnamefont {J.~D.}\
  \bibnamefont {Ziegler}}, \bibinfo {author} {\bibfnamefont {M.}~\bibnamefont
  {Selig}}, \bibinfo {author} {\bibfnamefont {T.}~\bibnamefont {Taniguchi}},
  \bibinfo {author} {\bibfnamefont {K.}~\bibnamefont {Watanabe}}, \bibinfo
  {author} {\bibfnamefont {A.}~\bibnamefont {Knorr}}, \bibinfo {author}
  {\bibfnamefont {A.}~\bibnamefont {Chernikov}},\ and\ \bibinfo {author}
  {\bibfnamefont {E.}~\bibnamefont {Malic}},\ }\bibfield  {title} {\bibinfo
  {title} {{Temporal Evolution of Low-Temperature Phonon Sidebands in
  Transition Metal Dichalcogenides}},\ }\href
  {https://doi.org/10.1021/acsphotonics.0c00866} {\bibfield  {journal}
  {\bibinfo  {journal} {ACS Photonics}\ }\textbf {\bibinfo {volume} {7}},\
  \bibinfo {pages} {2756} (\bibinfo {year} {2020})}\BibitemShut {NoStop}%
\bibitem [{\citenamefont {Selig}\ \emph {et~al.}(2016)\citenamefont {Selig},
  \citenamefont {Bergh{\"{a}}user}, \citenamefont {Raja}, \citenamefont
  {Nagler}, \citenamefont {Sch{\"{u}}ller}, \citenamefont {Heinz},
  \citenamefont {Korn}, \citenamefont {Chernikov}, \citenamefont {Malic},\ and\
  \citenamefont {Knorr}}]{Selig2016}%
  \BibitemOpen
  \bibfield  {author} {\bibinfo {author} {\bibfnamefont {M.}~\bibnamefont
  {Selig}}, \bibinfo {author} {\bibfnamefont {G.}~\bibnamefont
  {Bergh{\"{a}}user}}, \bibinfo {author} {\bibfnamefont {A.}~\bibnamefont
  {Raja}}, \bibinfo {author} {\bibfnamefont {P.}~\bibnamefont {Nagler}},
  \bibinfo {author} {\bibfnamefont {C.}~\bibnamefont {Sch{\"{u}}ller}},
  \bibinfo {author} {\bibfnamefont {T.~F.}\ \bibnamefont {Heinz}}, \bibinfo
  {author} {\bibfnamefont {T.}~\bibnamefont {Korn}}, \bibinfo {author}
  {\bibfnamefont {A.}~\bibnamefont {Chernikov}}, \bibinfo {author}
  {\bibfnamefont {E.}~\bibnamefont {Malic}},\ and\ \bibinfo {author}
  {\bibfnamefont {A.}~\bibnamefont {Knorr}},\ }\bibfield  {title} {\bibinfo
  {title} {{Excitonic linewidth and coherence lifetime in monolayer transition
  metal dichalcogenides}},\ }\href {https://doi.org/10.1038/ncomms13279}
  {\bibfield  {journal} {\bibinfo  {journal} {Nat. Commun.}\ }\textbf {\bibinfo
  {volume} {7}},\ \bibinfo {pages} {13279} (\bibinfo {year}
  {2016})}\BibitemShut {NoStop}%
\bibitem [{\citenamefont {Carbone}\ \emph {et~al.}(2020)\citenamefont
  {Carbone}, \citenamefont {Mayers},\ and\ \citenamefont
  {Reichman}}]{Carbone2020}%
  \BibitemOpen
  \bibfield  {author} {\bibinfo {author} {\bibfnamefont {M.~R.}\ \bibnamefont
  {Carbone}}, \bibinfo {author} {\bibfnamefont {M.~Z.}\ \bibnamefont
  {Mayers}},\ and\ \bibinfo {author} {\bibfnamefont {D.~R.}\ \bibnamefont
  {Reichman}},\ }\bibfield  {title} {\bibinfo {title} {{Microscopic model of
  the doping dependence of linewidths in monolayer transition metal
  dichalcogenides}},\ }\href {https://doi.org/10.1063/5.0008730} {\bibfield
  {journal} {\bibinfo  {journal} {J. Chem. Phys.}\ }\textbf {\bibinfo {volume}
  {152}},\ \bibinfo {pages} {194705} (\bibinfo {year} {2020})}\BibitemShut
  {NoStop}%
\bibitem [{\citenamefont {Leo}\ \emph {et~al.}(1988)\citenamefont {Leo},
  \citenamefont {R{\"{u}}hle},\ and\ \citenamefont {Ploog}}]{Leo1988}%
  \BibitemOpen
  \bibfield  {author} {\bibinfo {author} {\bibfnamefont {K.}~\bibnamefont
  {Leo}}, \bibinfo {author} {\bibfnamefont {W.~W.}\ \bibnamefont
  {R{\"{u}}hle}},\ and\ \bibinfo {author} {\bibfnamefont {K.}~\bibnamefont
  {Ploog}},\ }\bibfield  {title} {\bibinfo {title} {{Hot-carrier energy-loss
  rates in GaAs/AlxGa1-xAs quantum wells}},\ }\href
  {https://doi.org/10.1103/PhysRevB.38.1947} {\bibfield  {journal} {\bibinfo
  {journal} {Phys. Rev. B}\ }\textbf {\bibinfo {volume} {38}},\ \bibinfo
  {pages} {1947} (\bibinfo {year} {1988})}\BibitemShut {NoStop}%
\bibitem [{\citenamefont {Mak}\ \emph {et~al.}(2013)\citenamefont {Mak},
  \citenamefont {He}, \citenamefont {Lee}, \citenamefont {Lee}, \citenamefont
  {Hone}, \citenamefont {Heinz},\ and\ \citenamefont {Shan}}]{Mak2012}%
  \BibitemOpen
  \bibfield  {author} {\bibinfo {author} {\bibfnamefont {K.~F.}\ \bibnamefont
  {Mak}}, \bibinfo {author} {\bibfnamefont {K.}~\bibnamefont {He}}, \bibinfo
  {author} {\bibfnamefont {C.}~\bibnamefont {Lee}}, \bibinfo {author}
  {\bibfnamefont {G.~H.}\ \bibnamefont {Lee}}, \bibinfo {author} {\bibfnamefont
  {J.}~\bibnamefont {Hone}}, \bibinfo {author} {\bibfnamefont {T.~F.}\
  \bibnamefont {Heinz}},\ and\ \bibinfo {author} {\bibfnamefont
  {J.}~\bibnamefont {Shan}},\ }\bibfield  {title} {\bibinfo {title} {{Tightly
  bound trions in monolayer MoS$_2$.}},\ }\href
  {https://doi.org/10.1038/nmat3505} {\bibfield  {journal} {\bibinfo  {journal}
  {Nat. Mater.}\ }\textbf {\bibinfo {volume} {12}},\ \bibinfo {pages} {207}
  (\bibinfo {year} {2013})}\BibitemShut {NoStop}%
\bibitem [{\citenamefont {Korm{\'{a}}nyos}\ \emph {et~al.}(2015)\citenamefont
  {Korm{\'{a}}nyos}, \citenamefont {Burkard}, \citenamefont {Gmitra},
  \citenamefont {Fabian}, \citenamefont {Z{\'{o}}lyomi}, \citenamefont
  {Drummond},\ and\ \citenamefont {Fal'ko}}]{Kormanyos2015}%
  \BibitemOpen
  \bibfield  {author} {\bibinfo {author} {\bibfnamefont {A.}~\bibnamefont
  {Korm{\'{a}}nyos}}, \bibinfo {author} {\bibfnamefont {G.}~\bibnamefont
  {Burkard}}, \bibinfo {author} {\bibfnamefont {M.}~\bibnamefont {Gmitra}},
  \bibinfo {author} {\bibfnamefont {J.}~\bibnamefont {Fabian}}, \bibinfo
  {author} {\bibfnamefont {V.}~\bibnamefont {Z{\'{o}}lyomi}}, \bibinfo {author}
  {\bibfnamefont {N.~D.}\ \bibnamefont {Drummond}},\ and\ \bibinfo {author}
  {\bibfnamefont {V.}~\bibnamefont {Fal'ko}},\ }\bibfield  {title} {\bibinfo
  {title} {{K {\textperiodcentered} P Theory for Two-Dimensional Transition
  Metal Dichalcogenide Semiconductors}},\ }\href
  {https://doi.org/10.1088/2053-1583/2/2/022001} {\bibfield  {journal}
  {\bibinfo  {journal} {2D Mater.}\ }\textbf {\bibinfo {volume} {2}},\ \bibinfo
  {pages} {022001} (\bibinfo {year} {2015})}\BibitemShut {NoStop}%
\bibitem [{\citenamefont {Larentis}\ \emph {et~al.}(2018)\citenamefont
  {Larentis}, \citenamefont {Movva}, \citenamefont {Fallahazad}, \citenamefont
  {Kim}, \citenamefont {Behroozi}, \citenamefont {Taniguchi}, \citenamefont
  {Watanabe}, \citenamefont {Banerjee},\ and\ \citenamefont
  {Tutuc}}]{Larentis2018}%
  \BibitemOpen
  \bibfield  {author} {\bibinfo {author} {\bibfnamefont {S.}~\bibnamefont
  {Larentis}}, \bibinfo {author} {\bibfnamefont {H.~C.~P.}\ \bibnamefont
  {Movva}}, \bibinfo {author} {\bibfnamefont {B.}~\bibnamefont {Fallahazad}},
  \bibinfo {author} {\bibfnamefont {K.}~\bibnamefont {Kim}}, \bibinfo {author}
  {\bibfnamefont {A.}~\bibnamefont {Behroozi}}, \bibinfo {author}
  {\bibfnamefont {T.}~\bibnamefont {Taniguchi}}, \bibinfo {author}
  {\bibfnamefont {K.}~\bibnamefont {Watanabe}}, \bibinfo {author}
  {\bibfnamefont {S.~K.}\ \bibnamefont {Banerjee}},\ and\ \bibinfo {author}
  {\bibfnamefont {E.}~\bibnamefont {Tutuc}},\ }\bibfield  {title} {\bibinfo
  {title} {Large effective mass and interaction-enhanced zeeman splitting of
  $k$-valley electrons in ${\mathrm{mose}}_{2}$},\ }\href
  {https://doi.org/10.1103/PhysRevB.97.201407} {\bibfield  {journal} {\bibinfo
  {journal} {Phys. Rev. B}\ }\textbf {\bibinfo {volume} {97}},\ \bibinfo
  {pages} {201407} (\bibinfo {year} {2018})}\BibitemShut {NoStop}%
\bibitem [{\citenamefont {Wagner}\ \emph
  {et~al.}(2020{\natexlab{b}})\citenamefont {Wagner}, \citenamefont {Wietek},
  \citenamefont {Ziegler}, \citenamefont {Semina}, \citenamefont {Taniguchi},
  \citenamefont {Watanabe}, \citenamefont {Zipfel}, \citenamefont {Glazov},\
  and\ \citenamefont {Chernikov}}]{PhysRevLett.125.267401}%
  \BibitemOpen
  \bibfield  {author} {\bibinfo {author} {\bibfnamefont {K.}~\bibnamefont
  {Wagner}}, \bibinfo {author} {\bibfnamefont {E.}~\bibnamefont {Wietek}},
  \bibinfo {author} {\bibfnamefont {J.~D.}\ \bibnamefont {Ziegler}}, \bibinfo
  {author} {\bibfnamefont {M.~A.}\ \bibnamefont {Semina}}, \bibinfo {author}
  {\bibfnamefont {T.}~\bibnamefont {Taniguchi}}, \bibinfo {author}
  {\bibfnamefont {K.}~\bibnamefont {Watanabe}}, \bibinfo {author}
  {\bibfnamefont {J.}~\bibnamefont {Zipfel}}, \bibinfo {author} {\bibfnamefont
  {M.~M.}\ \bibnamefont {Glazov}},\ and\ \bibinfo {author} {\bibfnamefont
  {A.}~\bibnamefont {Chernikov}},\ }\bibfield  {title} {\bibinfo {title}
  {Autoionization and dressing of excited excitons by free carriers in
  monolayer ${\mathrm{wse}}_{2}$},\ }\href
  {https://doi.org/10.1103/PhysRevLett.125.267401} {\bibfield  {journal}
  {\bibinfo  {journal} {Phys. Rev. Lett.}\ }\textbf {\bibinfo {volume} {125}},\
  \bibinfo {pages} {267401} (\bibinfo {year} {2020}{\natexlab{b}})}\BibitemShut
  {NoStop}%
\bibitem [{\citenamefont {Chevy}(2006)}]{PhysRevA.74.063628}%
  \BibitemOpen
  \bibfield  {author} {\bibinfo {author} {\bibfnamefont {F.}~\bibnamefont
  {Chevy}},\ }\bibfield  {title} {\bibinfo {title} {Universal phase diagram of
  a strongly interacting fermi gas with unbalanced spin populations},\ }\href
  {https://doi.org/10.1103/PhysRevA.74.063628} {\bibfield  {journal} {\bibinfo
  {journal} {Phys. Rev. A}\ }\textbf {\bibinfo {volume} {74}},\ \bibinfo
  {pages} {063628} (\bibinfo {year} {2006})}\BibitemShut {NoStop}%
\bibitem [{\citenamefont {Wasak}\ \emph {et~al.}(2021)\citenamefont {Wasak},
  \citenamefont {Pientka},\ and\ \citenamefont {Piazza}}]{Wasak2021}%
  \BibitemOpen
  \bibfield  {author} {\bibinfo {author} {\bibfnamefont {T.}~\bibnamefont
  {Wasak}}, \bibinfo {author} {\bibfnamefont {F.}~\bibnamefont {Pientka}},\
  and\ \bibinfo {author} {\bibfnamefont {F.}~\bibnamefont {Piazza}},\
  }\bibfield  {title} {\bibinfo {title} {{Fermi polaron laser in
  two-dimensional semiconductors}},\ }\href {http://arxiv.org/abs/2103.14040}
  {\  (\bibinfo {year} {2021})},\ \Eprint {https://arxiv.org/abs/2103.14040}
  {arXiv:2103.14040} \BibitemShut {NoStop}%
\end{thebibliography}

%

\newpage

\appendix

\section{Pump power dependence of trion/Fermi-polaron cooling}
\label{app:Density}

In this section we present the dependence of the measured cooling dynamics on the pump power of the excitation laser.
The experimental approach to determine transient trion/Fermi-polaron temperatures is described in Sect.\,\ref{cooling-temp:exp} of the main manuscript.
It is based on the time-resolved analysis of the low-energy recoil flank. 
For the measurements presented here, the lattice temperature was set to 5\,K and the free hole density was fixed to about $n_h=1\times10^{11}$\,cm$^{-2}$. 
The effect of the photo-excitation on the free carrier doping was compensated by readjusting the gate voltage during the measurement (see Appendix \ref{app:Doping}).
The injected electron-hole pair density is obtained as $n_{eh}=4\alpha P/(E_{ph} f_{rep} \pi w^2)$, with the pump power P, repetition rate $f_{rep}=80$\,MHz, full-width-at-half-maximum of the spot of $w=1 \mu$m,  excitation energy $E_{ph}=1.663$\,eV and an effective absorption of $\alpha=2\%$.
The latter is estimated from absorption type reflectance measurements.

\begin{figure}[ht]
	\centering
			\includegraphics[width=6 cm]{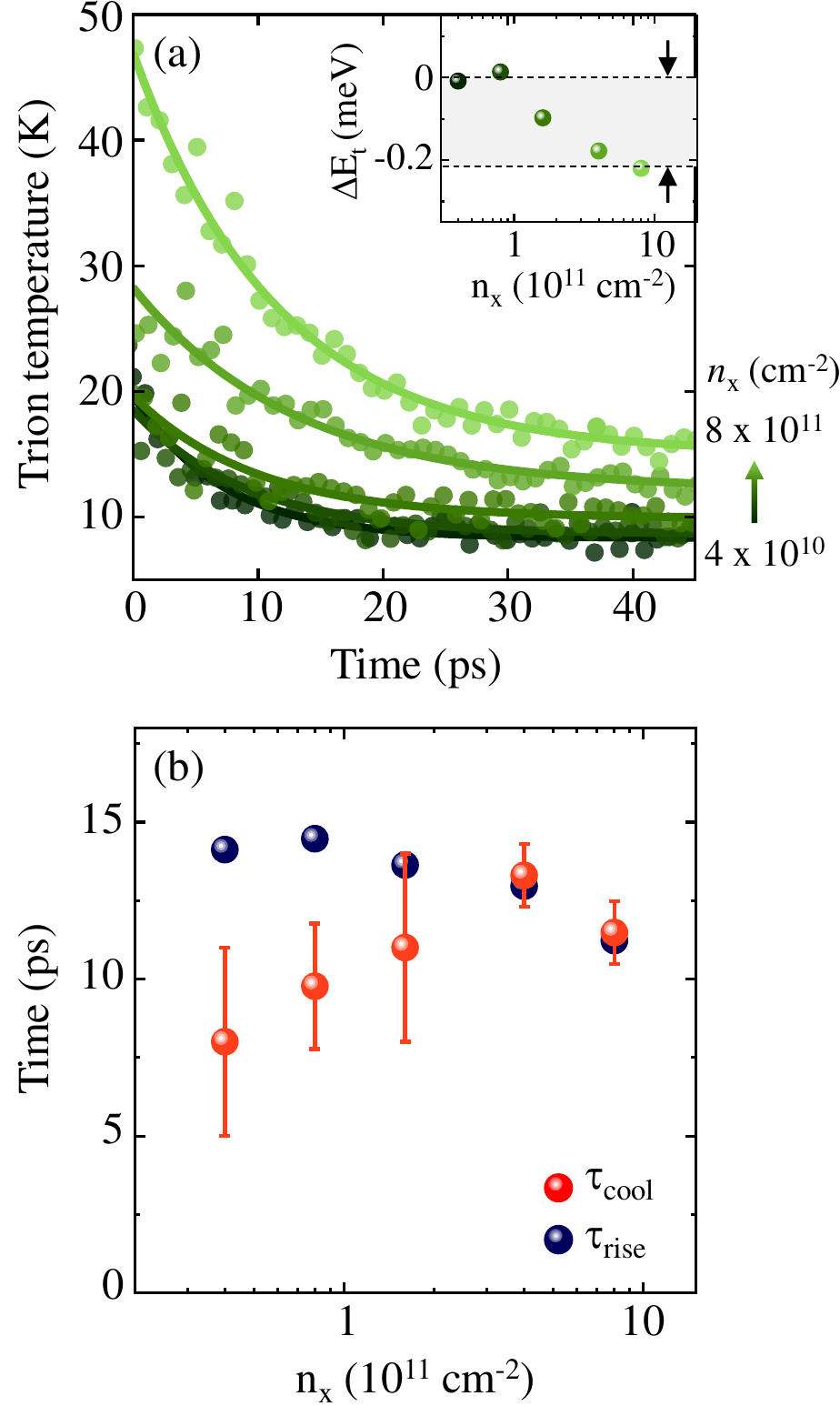}
		\caption{ 
		(a) Transient trion temperature for different injected electron-hole pair densities $n_{eh}$ at a fixed hole density of $n_h \approx 1\times10^{11}$ cm$^{-2}$, near-resonant excitation at $1.663$ eV and lattice temperature T $= 5$ K. 
		Solid lines correspond to mono-exponential fits of the data. 
		The extrapolated zero-density trion temperature after 50\,ps is 7.6\,K. 
		The inset shows the corresponding X+ peak energy shift $\Delta$E$_0^{tr}$ as a function of pump density. 
		(b) Extracted 1/e cooling times from (a) as function of pump density $n_{eh}$. 
		Rise time of the trion PL is shown for comparison.
		}
	\label{figADensity}
\end{figure}

Fig.\,\ref{figADensity}\,(a) shows the transient trion temperature for several different pump densities $n_{eh}$ in the range of $4\times10^{10}$ to $8\times10^{11}$ cm$^{-2}$.
We obtain similar cooling transients for the lowest pump densities, indicating linear excitation regime.
At these conditions, the extracted effective temperature after 50\,ps reaches 8\,K for the lowest pump density.
This value is very close to the nominal heat sink temperature of 5\,K.

At higher electron-hole pair densities, however, both excess and equilibrium temperatures increase, while the cooling time remains almost unchanged within experimental error, as illustrated in Fig.\,\ref{figADensity}\,(b). 
The observed increase of the excess and equilibrium temperatures is consistent with the observed shift of the trion emission peak to lower energies, shown in the inset of Fig.\,\ref{figADensity}\,(a). 
We also note that the shift of the neutral exciton peak energy is almost identical (not shown here). 
Transient increase of the trion/Fermi-polaron temperature may be associated with additional heating due to reabsorption of non-equilibrium phonons, created during initial relaxation, known as ``hot-phonon effect'' in the literature\,\cite{Leo1988}.
In addition, laser-induced heating of the lattice could contribute to the increase of the equilibrium temperature\,\cite{Park2021}.
The latter is consistent with the shift of all resonances to lower energies following temperature-induced decrease of the bandgap. 

We note that in the temperature-dependent experiments presented in Fig.\,\ref{fig1} and Fig.\,\ref{fig3} of the main manuscript we used an excitation density of $4$x$10^{11}$ cm$^{-2}$ for increased signal-to-noise ratio.
We would thus expect additional contributions from laser heating leading to
slightly larger absolute values of the extracted trion/Fermi-polaron temperatures.
These should be particularly pronounced for measurements at nominal heat sink temperatures T $\leq15$ K. 
Most importantly, however, the cooling times are robust with respect to the pump power.
As demonstrated in the main manuscript, they also match the values of the PL rise time with small deviations at lowest pump densities.

\section{Trion / Fermi-polaron cooling at different excitation energies}
\label{app:OffRes}

\begin{figure}[ht]
	\centering
			\includegraphics[width=5.5 cm]{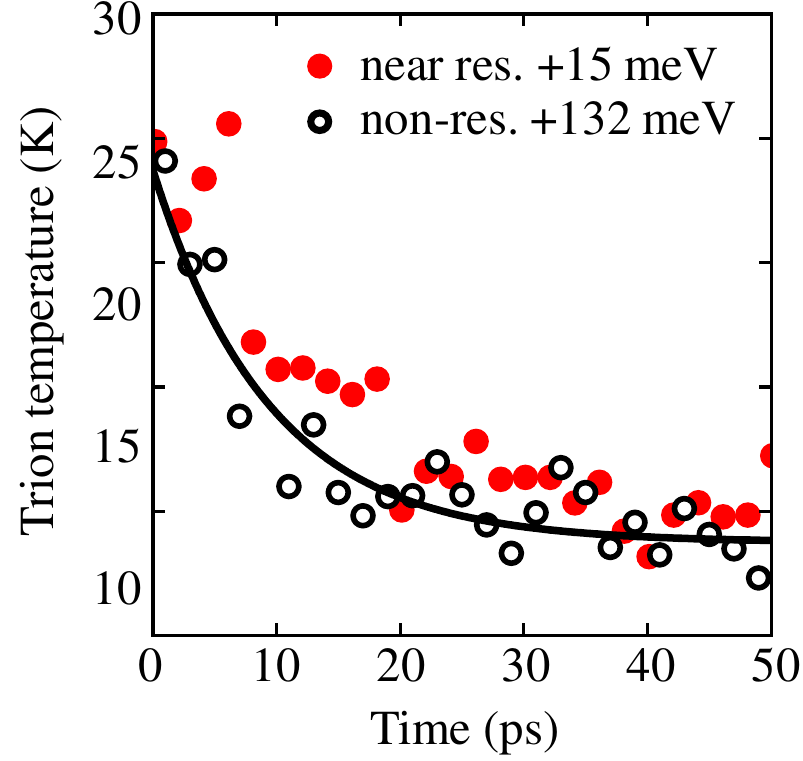}
		\caption{ 
		 Transient trion / Fermi-polaron temperature for near-resonant and non-resonant excitation of the neutral exciton transition. 
		 The pump density is set to a few $10^{11}$ cm$^{-2}$ and the hole density to $n_h \approx 1\times10^{11}$ cm$^{-2}$. 
		 Mono-exponential fit (solid line) corresponds to a cooling time of 9.5\,ps. 
		}
	\label{figAOffRes}
\end{figure}

The PL studies reported in the main manuscript are performed using pulsed excitation at the photon energy of 1.657\,eV, near-resonant to the A-exciton transition at $1.642$ eV. 
In this section, we present the comparison of these results with the case of non-resonant excitation with about 0.13\,eV excess energy by tuning the laser to 1.774\,eV. 
The lattice temperature is set to 5\,K and we consider the regime of low and comparable densities of the free holes and optically injected electron-hole pairs in the range of $10^{11}$\,cm$^{-2}$. 
As illustrated in Fig.\,\ref{figAOffRes}, we find essentially the same behavior of the transient trion / Fermi-polaron temperature for near-resonant and non-resonant excitation conditions. 
Both the initial temperatures and the cooling times are the same. 
This result strongly suggests that the detected dynamics are not determined by the initial relaxation of the optically injected electron-hole pairs.
The latter should thus occur on much faster, sub-picosecond time-scales below the resolution of the setup.

\section{Trion / Fermi-polaron cooling in the n-doped regime}
\label{app:nDoping}

As we illustrate in Figs.\,\ref{fig1}\,(a) and (b) both n- and p-type trions / attractive Fermi-polarons exhibit similar low-energy flanks in their PL due to the recoil effect.
For the time-resolved studies, however, discussed in the main manuscript we focus on the p-doped region due to the overall smaller symmetric broadening allowing for more accurate quantitative analysis.
In this section, we address the n-doped regime for direct comparison. 

\begin{figure}[ht]
	\centering
			\includegraphics[width=5 cm]{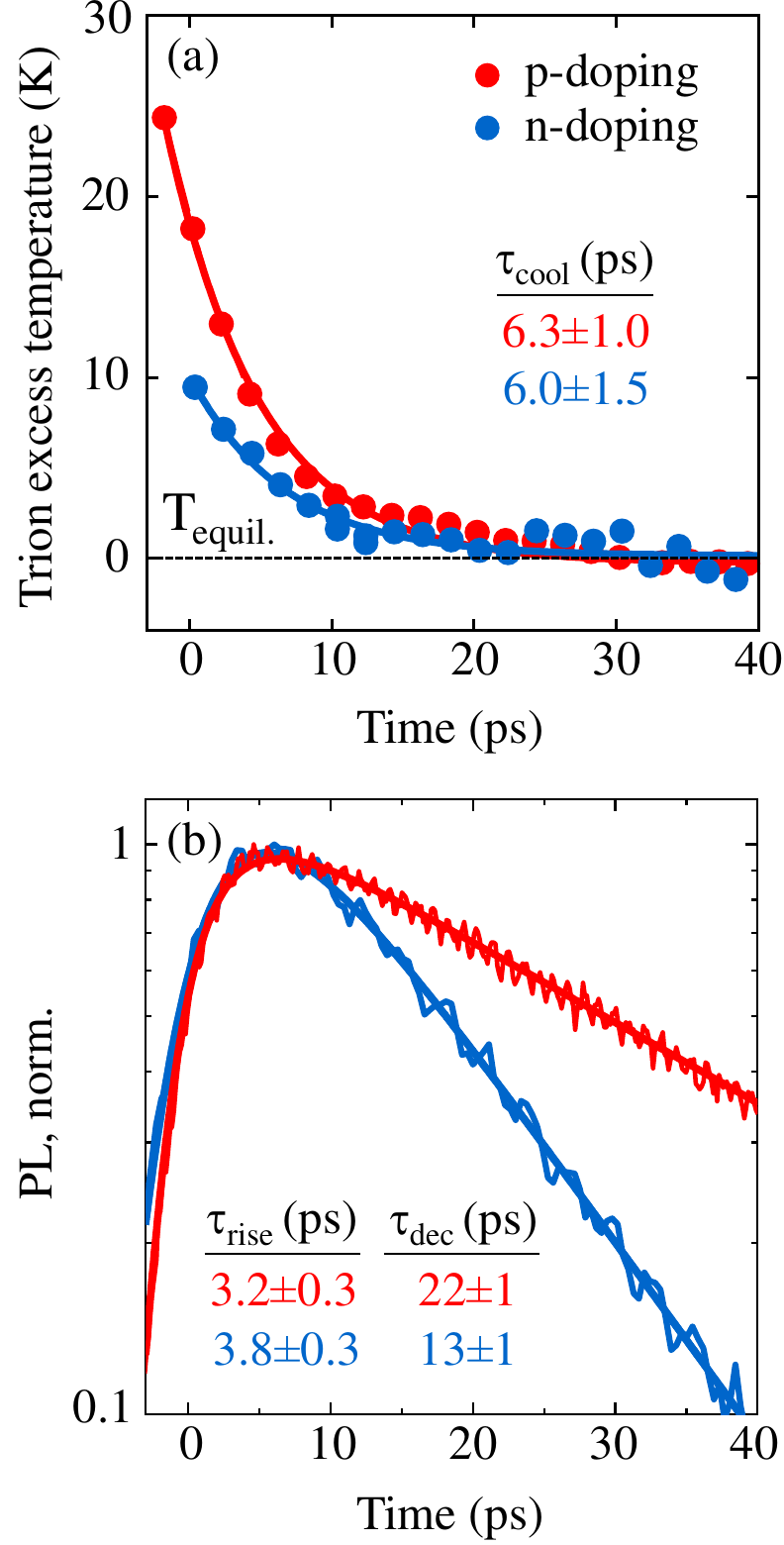}
		\caption{ 
		 (a) Transient excess trion temperature for estimated free carrier densities of $n_e \approx n_h \approx 7\times10^{11}$ cm$^{-2}$.  
		 The excess temperature is plotted with respect to the equilibrium temperature at 40\,ps. 
		 (b) Corresponding normalized trion PL transients in the p- and n- doped regime. 
		 Extracted exponential cooling times as well as PL rise and decay times are illustrated. 
		 }
	\label{figAnDoping}
\end{figure}

In Fig.\,\ref{figAnDoping}\,(a) we compare the extracted transient trion temperature at free carrier densities $n_e \approx n_h \approx 7\times10^{11}$cm$^{-2}$ at the lattice temperature of 5\,K. 
Here, we follow the same procedure as discussed in the main manuscript the trion temperature.
We further consider the light-matter coupling exponent $\epsilon^*$ to be same for the p- and n-doped species due to matching binding energies. 
Both $X^+$ and $X^-$ features exhibit very similar initial excess temperatures on the order of 10 to 20\,K and, more importantly, almost equivalent cooling times.
The latter roughly correspond to the rise times of the PL intensity transients, presented in Fig.\,\ref{figAnDoping}\,(b).
We note, that only the decay times of the trions / Fermi-polarons are substantially shorter in the n-doped regime.
Overall, these results demonstrate that the cooling dynamics do not depend strongly on the charge of the free carriers.  

\section{Estimation of free charge carrier concentration}
\label{app:Doping}

In the studied devices the free charge carrier density is controlled by applying a variable gate voltage in a plate capacitor geometry: 
thin-layer graphite flake and MoSe$_2$ monolayer separated by an thin hBN spacer (additional details regarding fabrication of the devices are presented in the supplementary of Ref.\,\cite{Wagner2020}). 
While a capacitor model can be used to estimate free carrier density, it tends to be less reliable in the low-doping regime motivating the use of established spectroscopic observables instead.

\begin{figure}[t]
	\centering
			\includegraphics[width=5.5 cm]{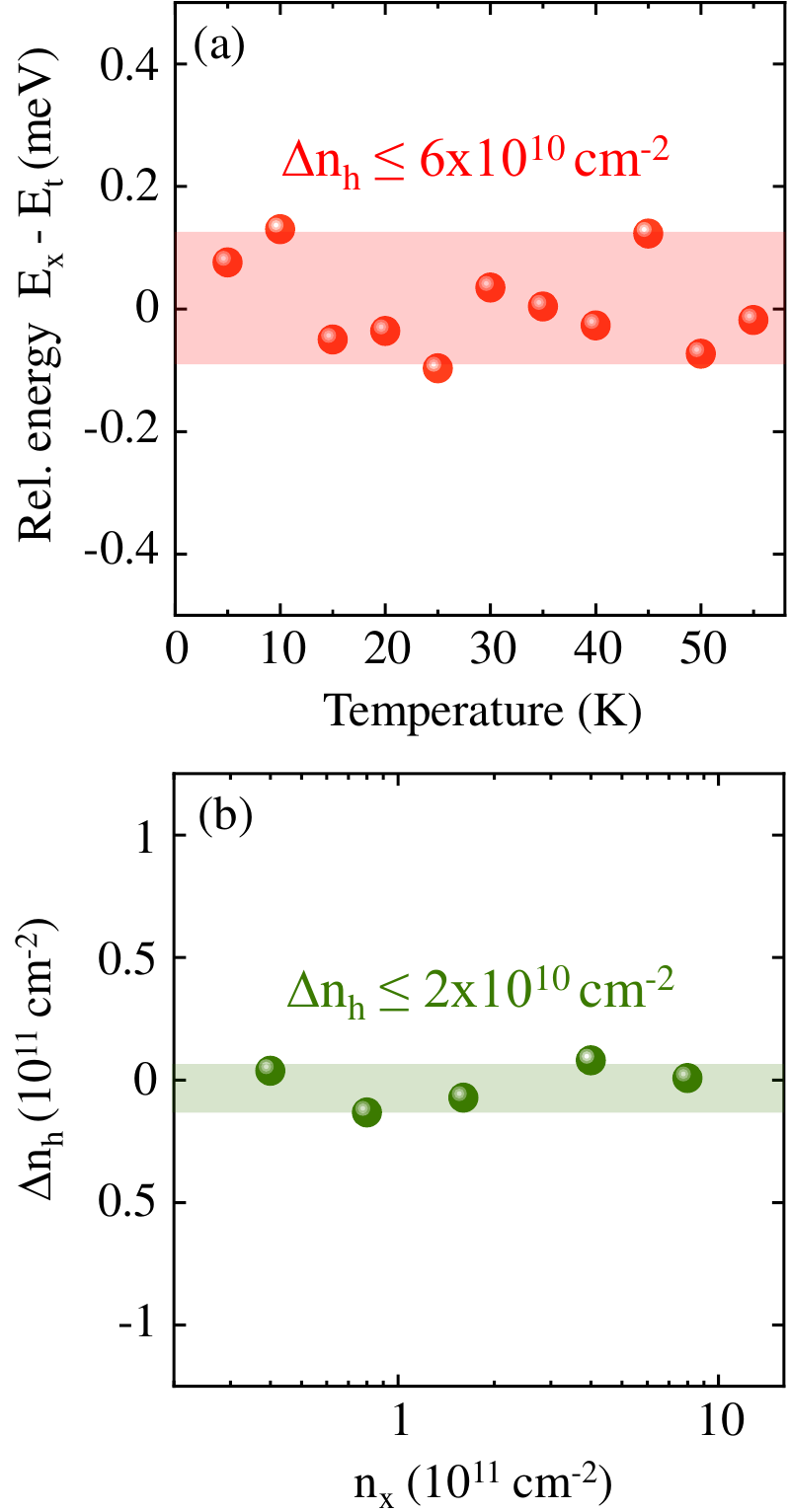}
		\caption{ 
		(a) Relative exciton-trion energy separation as function of temperature corresponding to the measurements presented in Fig.\,\ref{fig1}\,(c).
		(b) Relative exciton-trion energy separation as function of pump density $n_{eh}$ corresponding to the measurements presented in Fig.\,\ref{figADensity}. 
		The remaining experimental fluctuation of the doping level is estimated by setting the exciton-trion energy variations equal to the Fermi energy shift using a hole mass of $0.6$\,m$_0$. 
		In both (a) and (b) the average hole density is $1\times10^{11}$ cm$^{.2}$.
		}
	\label{figADopingCompensation}
\end{figure} 
In our study we use the relative energy separation between exciton and trion peaks (or repulsive and attractive Fermi polarons).
This quantity scales linearly with the free carrier density in both trion and Fermi polaron formalisms\,\cite{Glazov2020c}.
While there are small differences in the prefactors\,\cite{Glazov2020c}, resulting in systematic deviations on the order of 30\%, we follow the approach in Ref.\,\cite{Mak2012} and set it equal to the Fermi energy shift $\Delta E_f$. 
The energy separation is extracted from fitting gate voltage-dependent PL spectra with a multi-Lorentzian model in the low-density regime.
From this analysis, we also determine the onset for zero-doping at -0.2 and 0.4\,V for p- and n-doping, respectively.

Considering that free charge carriers reside in the energetically lowest conduction and valence bands at the K/K' valleys, the doping level is calculated as $n_{e,h}=\Delta E_f m_{e,h}/(\hbar^2\pi)$ with the effective masses of either $m_e=0.5$\,m$_0$ or $m_h=0.6$\,m$_0$\,\cite{Kormanyos2015}. 
From the dependence of exciton-trion energy separation on gate voltage we obtain scaling factors for the charge carrier densities $2\times10^{11}$ cm$^{-2}/V$ and $4.8\times10^{11}$ cm$^{-2}/V$ for p- and n-doping, respectively.
We note, however, that the extracted ratio $m_h/m_x=0.44$ from the analysis of temperature-dependent measurements in Fig.\,\ref{fig1}\,(e) with Eq.\,\eqref{expon-fit} indicates either a lower hole or a higher electron effective mass.
The latter would be consistent with transport experiments at in monolayer MoSe$_2$ \cite{Larentis2018} that demonstrate an electron mass as large as 0.8\,m$_0$.
In that case, electron densities would be underestimated in our analysis, while the hole densities should remain largely correct. 


We also note that that the doping density can slowly change over time as well as under optical illumination.
To compensate for that, we manually controlled the gate voltage to ensure constant doping density throughout all temperature- and pump-density dependent experiments presented in Figs.\,\ref{fig1}, Fig.\,\ref{fig4} and Fig.\,\ref{figADensity}. 
The gate voltage was slightly adjusted such that the exciton-trion energy separation remained fixed within a small experimental uncertainty far below 1\,meV.
Fig.\,\ref{figADopingCompensation}\,(a) and (b) show the corresponding relative exciton-trion energy separations as function of temperature and injected electron-hole pair density $n_{eh}$, respectively. 
The data illustrates residual fluctuations of the free charge carrier on the order of several 10$^{10}$\,cm$^{-2}$.

\section{Basics of the Fermi-polaron approach}\label{app:basic}

We use a simplified approach to the description of the exciton interaction with the resident charge carriers presented in Ref.~\cite{Glazov2020c}. For definiteness we consider an electron-doped sample (the modification of the results for the case of the hole doping are evident) and, for simplicity, we assume that the electrons are degenerate. We consider the regime where 
\begin{equation}
\label{req}
E_F \ll E_{b,tr} \ll E_{b,x}.
\end{equation}
Here $E_F$ is the electron Fermi energy, $E_{b,tr}$ is the trion binding energy and $E_{b,x}$ is the exciton binding energy, respectively. The exciton Greens function in the presence of the resident electrons is given by
\begin{multline}
\label{Gx:dressed}
\mathcal G_x(\varepsilon,\bm k) = \frac{1}{\varepsilon - E^x_{\bm k} - \Sigma_x(\varepsilon,\bm k)} \\
\approx \frac{Z_{\bm k}}{\varepsilon - E^{tr}_{\bm k}-N_e\mathcal D^{-1}}  + \frac{1-Z_{\bm k}}{\varepsilon - E^{x}_{\bm k}}. 
\end{multline}
The approximate equality holds for the energies in the vicinity of the trion (attractive Fermi-polaron) and exciton (repulsive Fermi-polaron) resonances.
Here $E^x_{\bm k}$ is the exciton energy dispersion, $E^{tr}_{\bm k}$ is the trion energy dispersion, $\Sigma_x(\varepsilon,\bm k)$ is the exciton self-energy, 
\begin{equation}
\label{Zk}
Z_{\bm k} \approx N_e /(\mathcal D E_{b,tr}),
\end{equation} is the renormalization factor proportional to the electron density $N_e$ which describes, in particular, the transfer of the oscillator strength between the repulsive polaron (exciton) and attractive polaron (trion). 

Generally, in agreement with~\cite{Suris2001,suris:correlation,PhysRevA.85.021602,Cotlet2019,Glazov2020c,PhysRevLett.125.267401} the exciton self-energy and the electron-exciton scattering amplitude obey the set of the self-consistent equations
\begin{subequations}
\label{sigma:self}
\begin{multline}
\Sigma_x(\varepsilon,\bm k) = \int \frac{d\xi}{2\pi \mathrm i} \sum_{\bm p} G_e(\xi,\bm p)  T\left(\varepsilon {+} \xi, \bm k + \bm p \right) \\
= \sum_{\bm p} n_{{\bm p}} T\left(\varepsilon + \frac{\hbar^2 p^2}{2m_e} , \bm k + \bm p \right),\label{sigma:self:A}
\end{multline}
\begin{multline}
1/T(\varepsilon,\bm k) = -\mathcal D \ln{\left(\frac{\bar E}{E_{b,tr}}\right)}  \\
{+} \sum_{\bm p} (1-n_{\bm p})\mathcal G_x\left(\varepsilon - \frac{\hbar^2 p^2}{2m_e} ;\bm k -\bm p\right),\label{T:gen}
\end{multline}
\end{subequations}
where $G_e(\xi,\bm p)$ is the electron Greens function, $n_{\bm p}$ is the electron distribution function, $T$ is the scattering amplitude, $\mathcal D$ is the exciton-electron reduced density of states, $\bar E$ is the cut-off energy and $E_{b,tr}>0$ is the trion binding energy. Hereafter we set the normalization area to unity. At negligible electron density where Eq.~\eqref{req} holds electron occupation numbers in Eq.~\eqref{T:gen}  can be set to zero.

In the non-self-consistent approach where in Eq.~\eqref{T:gen}  $\mathcal G_x$ is replaced by the bare exciton Greens function $G_x$ and $n_{\bm p}$ is neglected, we arrive at the second line of Eq.~\eqref{Gx:dressed} with $Z_{\bm k}$ given by Eq.~\eqref{Zk}, see Ref.~\cite{Glazov2020c} for details. Such an approach is equivalent to the Chevy representation of the Fermi-polaron state. Following Refs.~\cite{PhysRevA.74.063628,Sidler2016}, see also \cite{suris:correlation}, we present the wavefunction of the system ``exciton+Fermi-gas'' in the form
\begin{equation}
\label{chevy}
\Psi_{\bm k} = \varphi(\bm k) X_{\bm k}^\dag |0\rangle + \sum_{\bm p,\bm q} F_{\bm p,\bm q}(\bm k) X^\dag_{\bm k+\bm q-\bm p} e^\dag_{\bm p} e_{\bm q} |0\rangle.
\end{equation}
Here $|0\rangle$ is the ground state of the (doped) semiconductor with the filled valence band and partially filled (up to $k=k_F$) conduction band, $e^\dag$, $e$ are the conduction electron creation and annihilation operators, $X^\dag$, $X$ is the exciton creation and annihilation operators. In Eq.~\eqref{chevy} the summation over $\bm q$ is limited by $q< k_F$. At small doping, where the product to the trion radius, $a_{tr}$, and the Fermi wavevector is small, $a_{tr} k_F \ll 1$ [cf. Eq.~\eqref{req}], the dependence of $F_{\bm p,\bm q}$ on $\bm q$ can be disregarded. In this limit, the function $F_{\bm p,\bm q}(\bm k)$ has a meaning of the ``trion'' wavefunction, it accounts for the correlation between the exciton and electron. 
Under condition of Eq.~\eqref{req} and replacing the summation over $\bm q$ by the multiplication by the electron density $N_e$ we arrive at the following equations for the functions $\varphi(\bm k)$ and $F_{\bm p, \bm q}$ and the energy of the Fermi-polaron $\varepsilon$:
\begin{subequations}
\label{chevy:eqs}
\begin{multline}
E^x_{\bm k} \varphi(\bm k) + V_0 N_e \varphi(\bm k) \\
+ V_0N_e \sum_{\bm p} F_{\bm p,\bm q}(\bm k)=\varepsilon \varphi(\bm k),
\end{multline}
\begin{multline}
\left(E^x_{\bm k - \bm p} + E^e_{\bm p}\right)F_{\bm p,\bm q}(\bm k) \\
+ V_0 \varphi(\bm k) + V_0 \sum_{\bm p'} F_{\bm p',\bm q}(\bm k) = \varepsilon F_{\bm p, \bm q}(\bm k),
\end{multline}
\end{subequations}
where $V_0$ is the bare constant of the electron-exciton scattering. Assuming $V_0$ to be small and working in the linear order in $N_e$ we obtain
\begin{subequations}
\label{chevy:sol}
\begin{align}
&\varphi(\bm k) \approx \frac{V_0 N_e}{-E_{b,tr}} \mathcal F(\bm k),\\
&F_{\bm p, \bm q}(\bm k) \approx \frac{V_0}{E_{\bm k}^{tr} - E_{\bm k- \bm p}^x - E_{\bm p}^e}\mathcal F(\bm k),\label{chevy:sol:F}
\end{align}
\end{subequations}
where
\begin{equation}
\label{cond:chevy}
\mathcal F(\bm k) = \sum_{\bm p} F_{\bm p,\bm q}(\bm k).
\end{equation}
The self-consistency condition, Eq.~\eqref{cond:chevy}, yields the trion dispersion $E^{tr}_{\bm k} = E^{tr}_0 + \hbar^2 k^2/2m_{tr}$. The function $\mathcal F(\bm k)$ can be found from the normalization condition
\begin{equation}
\label{norm:chevy}
1 = |\varphi(\bm k)|^2 + \sum_{\bm p, \bm q} |F_{\bm p, \bm q}|^2 \approx \sum_{\bm p, \bm q} |F_{\bm p, \bm q}|^2 .
\end{equation}
At not too large $k$ ($\hbar^2 k^2/2\mu_{eX} \ll E_{b,tr}$) we find in the lowest order in $N_e$ that
\begin{equation}
\label{Fk:norm}
|\mathcal F(\bm k)|^2 \approx \frac{E_{b,tr}}{N_e \mathcal D V_0^2}.
\end{equation}

\section{Photoluminescence of Fermi-polarons}\label{app:FPPL}

In order to calculate the PL we evaluate the photon self-energy related to the photon generation as a result of the Fermi-polaron decay [Fig.~\ref{fig:K:PL}(a)]:
\begin{multline}
\label{sigma:PL:FP:Keld}
\Sigma_{PL}(\hbar\omega) = |\mathfrak M_r|^2 \\
\times
\int \frac{d\Omega}{2\pi\mathrm i} 
\int\frac{d\xi}{2\mathrm i\pi} \sum_{\bm q,\bm k} \mathcal G_x^{--}(\hbar\omega, 0)\mathcal G_x^{++}(\hbar\omega, 0) \\
\times
T^2(\hbar\omega+\hbar\Omega+\xi,\bm k + \bm q)
\mathcal G_x^{-+}(\hbar\omega+ \hbar\Omega,\bm k)\\
\times G_e^{-+}(\xi,\bm q) G_e^{+-}(\xi + \hbar\Omega,\bm q+ \bm k).
\end{multline}
Here $G_x(\varepsilon,\bm k)$ is the dressed exciton Keldysh Greens function, $G_e(\xi,\bm q)$ is the resident electron Greens function, superscripts $+$ and $-$ denote the branch of the Keldysh contour (cf. Ref. \cite{Glazov2020c}), and $T(\varepsilon, \bm k)$ is the exciton-electron scattering amplitude. Equation~\eqref{sigma:PL:FP:Keld} can be simplified taking into account the fact that the summation over $\bm q$ takes place over the occupied states: Hence, for degenerate carriers, $\xi$ and $E^e_{\bm q} \lesssim E_F$ that is much smaller than the trion binding energy. As a result, we can omit the dependence on $\bm q$ and $\xi$ in $T^2$.
As a result, we obtain
\begin{multline}
\label{sigma:PL:FP:Keld:1}
\Sigma_{PL}(\hbar\omega) = \frac{|\mathfrak M_r|^2}{(\hbar\omega - E_0^x)^2}
\int \frac{d\Omega}{2\pi\mathrm i} 
\sum_{\bm k}
T^2(\hbar\omega+\hbar\Omega,\bm k) \\
\times \Pi(\hbar\Omega, \bm k)
\mathcal G_x^{-+}(\hbar\omega+ \hbar\Omega,\bm k),
\end{multline}
with 
\begin{multline}
\label{Pi}
\Pi(\hbar\Omega,\bm k) = \sum_{\bm q} \int \frac{d\xi}{2\pi\mathrm i} G_e^{+-}(\xi+\hbar\Omega,\bm k+ \bm q) G_e^{-+}(\xi,\bm q) \\
\approx -2\pi \mathrm i N_e \delta(E^e_{\bm q} - \hbar\Omega) ,
\end{multline}
being the electron polarizability. The last approximate equality holds at $k\gg k_F$. This condition enables us to neglect the dynamics of the hole in the Fermi sea, cf. Ref.~\cite{suris:correlation}, and the Pauli blocking effects for the recoil electron.

Using Eq.~\eqref{Pi} (last approximate equality), assuming thermal equilibrium of Fermi-polarons with 
\begin{equation}
\label{equilibrium}
\mathcal G_x^{-+}(\varepsilon, \bm k) = 2\pi \mathrm i  f(\varepsilon) Z_{\bm k}\delta(\varepsilon - E_{\bm k}^{tr} - N_e/\mathcal D),
\end{equation} 
with quasi-equilibrium distribution $f(\varepsilon) \propto \exp{(-\varepsilon/k_B T_*)}$, cf. Eq.~\eqref{f:tr}, using the expression
\begin{equation}
\label{Tmat}
T(\varepsilon, \bm k) = \frac{\mathcal D^{-1} E_{b,tr}}{\varepsilon - E_{\bm k}^{tr} + \mathrm i \delta},
\end{equation}
valid in the vicinity of the attractive Fermi-polaron resonance, and retaining small energy shift $N_e/\mathcal D$ in the argument of $T$ only (which is crucial to avoid the divergence and properly account for the renormalization factors)  we arrive at Eq.~\eqref{PL:FP:Keld} of the main text.

\section{Radiative decay rate}\label{app:rad}

Self-energy is shown by Fig.~\ref{fig:K:PL}(b) and reads
\begin{multline}
\label{sigma:k:phot}
\Sigma_{phot}(\varepsilon, \bm k) 
=
 \int d\omega \sum_{\bm q,\bm p} \int \frac{d\xi}{2\pi\mathrm i} \int \frac{d\xi'}{2\pi\mathrm i} \\
 \times T(\varepsilon+\xi, \bm k+\bm q) G_e(\xi,\bm q) G_e(\xi',\bm p) T(\varepsilon+\xi ,\bm p)\\
 \times  \delta_{\bm p,\bm k+\bm q}|\mathfrak M_r|^2 \mathcal G_x^2(\varepsilon+\xi - \xi',0) G_{phot}(\varepsilon+\xi-\xi',\hbar\omega).
\end{multline} 
Here  $G_{phot}(\varepsilon,\hbar\omega) = \pi \mathrm i \mathcal D_{phot} \delta(\varepsilon- \hbar\omega)$ is the imaginary part of the photon Greens function, $\mathcal D_{phot}$ is the photon density of states at the energy $\varepsilon$. For simplicity we omitted the polarization-dependent factors in Eq.~\eqref{sigma:k:phot}. As in Sec.~\ref{app:FPPL} the summation over $\bm q$ is restricted by the occupied states, thus $\xi, E_{\bm q}^e \lesssim E_F$ are much smaller than other relevant energies, including $E_{b,tr}$ and $E^e_{\bm k}$, $E^{tr}_{\bm k}$. As a result, Eq.~\eqref{sigma:k:phot} simplifies to 
\begin{multline}
\label{sigma:k:phot:11}
\Sigma_{phot}(\varepsilon, \bm k)   
=
\int d\omega |\mathfrak M_r|^2 T^2(\varepsilon, \bm k)   \int \frac{d\xi'}{2\pi\mathrm i} \\
\times
 \Pi(\xi',\bm k) \mathcal G_x^2(\varepsilon - \xi',0) G_{phot}(\varepsilon-\xi').
\end{multline} 
The remaining integrations are trivial and yield Eq.~\eqref{sigma:k:phot:1} of the main text.

\section{Fermi-polaron cooling}\label{app:cool}

\begin{figure}[ht]
\includegraphics[width=\linewidth]{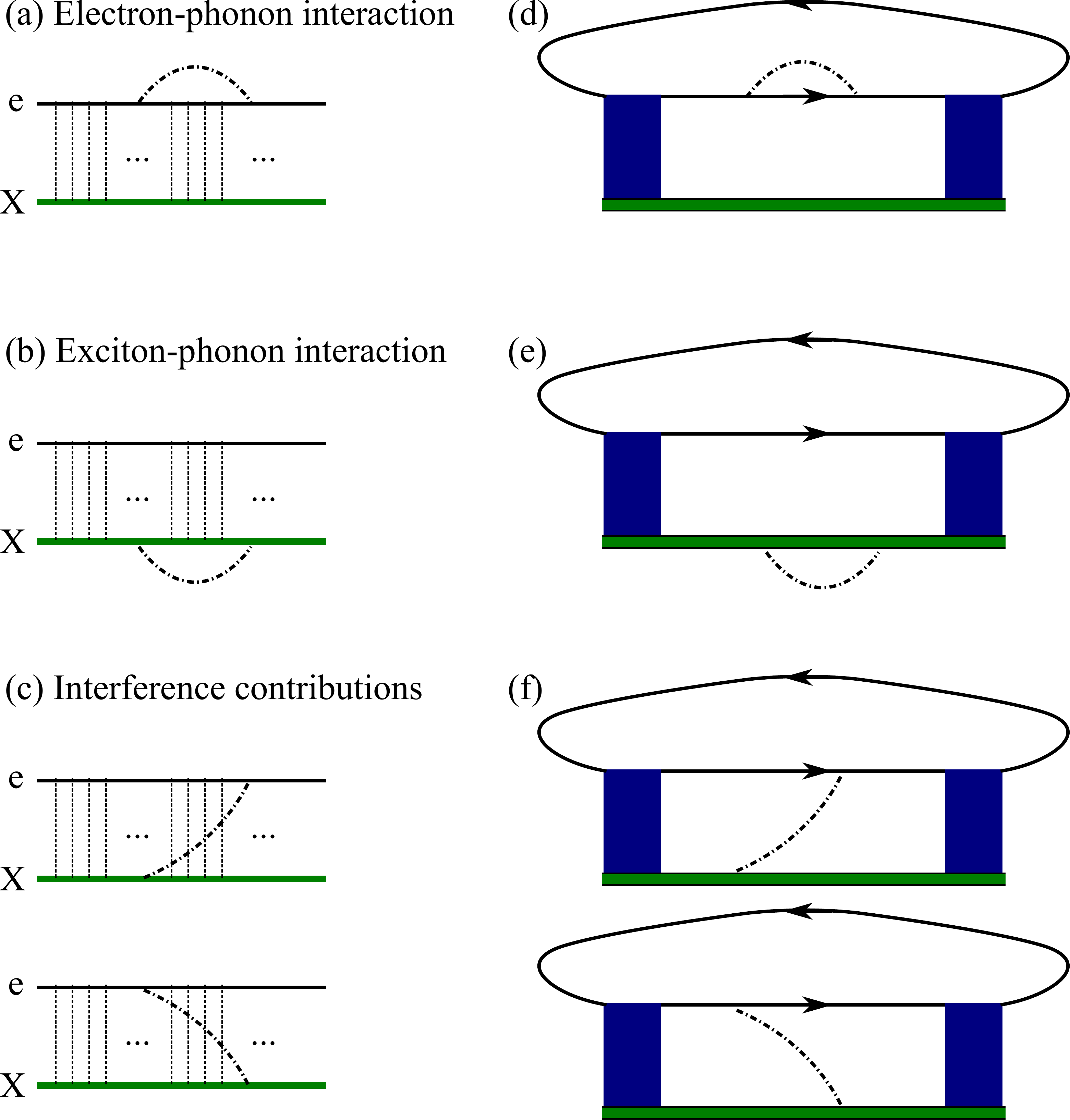}
\caption{Diagrams describing the Fermi-polaron interaction with phonons. We consider a $n$-type system with resident electrons. Dash-dotted line is the phonon Greens function. (a-c) Basic diagrams showing individual contributions resulting in $(\Xi_c - \Xi_v)^2$, $\Xi_c^2$ and $2\Xi_c(\Xi_c-\Xi_v)$ terms in $\mathcal B_{tr}(q)$, Eq.~\eqref{def:ac}. (d-f) Fermi-polaron self-energies. Note that the diagrams with at least one of the phonon vertices describing the Fermi sea hole interaction with phonons (not shown) are negligible at small electron densities. The diagrams with the phonon line encompassing $\mathcal G_x$ (not shown) have an additional smallness $\propto N_e/(\mathcal D E_{b,tr})$.}\label{fig:diag:cooling}
\end{figure}

Relevant diagrams describing Fermi-polaron coupling with phonons are shown in Fig.~\ref{fig:diag:cooling}. In what follows we consider the situation where the typical wavevector transferred in the course of the Fermi-polaron-phonon scattering exceeds the Fermi wavevector of the resident charge carriers $k_F$. This assumption allows us to neglect the interaction of the Fermi-sea hole with the phonon because the momentum of the Fermi-sea-hole is limited by $k_F$. As a result we are left with the essentially exciton- and electron-phonon interaction, or, in terms of constituent quasi-particles, the phonon interaction with the electron-in-exciton, outer electron in trion and hole-in-exciton.

The Keldysh technique allows to derive the kinetic equation for the quasi-particles distribution function. Correspondingly, the expressions for the cooling rate in the form of Eqs.~\eqref{Q:gen} and \eqref{rate:trions} can be derived, cf.~\cite{Glazov2020c,Wasak2021}. The key issue is to calculate the corresponding self-energies or matrix elements of the scattering. The relevant diagrams for the self energies are shown in Fig.~\ref{fig:diag:cooling}(d-e). It is illustrative, however, to derive the transition matrix elements
\begin{equation}
M_{\bm k\to \bm k'}^{FP,\pm \bm q} = \langle \Psi_{\bm k'} | V_{ph}^\pm(\bm q)|\langle \Psi_{\bm k}\rangle,
\end{equation}
 directly using the Chevy ansatz wavefunctions, Eq.~\eqref{chevy}. Here $V_{ph}^\pm$ are the operators of the electron (and exciton) interaction with longitudinal acoustic phonons~\cite{shree2018exciton}:
\begin{subequations}
\label{V:ph:pm}
\begin{multline}
V_{ph}^+(\bm q) = \sqrt{1+n^{ph}_{\bm q}} \sqrt{\frac{\hbar}{2\varrho \omega_{\bm q}^{ph}}} q \\
\times
\left[\sum_{\bm k} (\Xi_c - \Xi_v) X^{\dag}_{\bm k - \bm q} X_{\bm k} + \sum_{\bm p} \Xi_c  e^{\dag}_{\bm p - \bm q} e_{\bm p}\right],
\end{multline}
\begin{multline}
V_{ph}^-(\bm q) = \sqrt{n^{ph}_{\bm q}} \sqrt{\frac{\hbar}{2\varrho \omega_{\bm q}^{ph}}} q\\
\times 
 \left[\sum_{\bm k} (\Xi_c - \Xi_v) X^{\dag}_{\bm k + \bm q} X_{\bm k} + \sum_{\bm p} \Xi_c  e^{\dag}_{\bm p + \bm q} e_{\bm p}\right].
\end{multline}
\end{subequations}
Hre $\Xi_c$ and $\Xi_v$ are the deformation potential constants for the conduction and valence bands. The extension of the approach to account for other phonon modes and other types of interaction is straightforward.

In the lowest in the resident electron density $N_e$ order, the matrix elements $M_{\bm k\to \bm k'}^{FP,\pm \bm q}$ are contributed by the transitions with the excited Fermi sea [second term in Eq.~\eqref{chevy}]. For example,
\begin{multline}
\label{V:ph:m}
M_{\bm k\to \bm k'}^{FP,- \bm q} = \sqrt{n^{ph}_{\bm q}} \sqrt{\frac{\hbar}{2\varrho \omega_{\bm q}^{ph}}} q  \\
\times
\sum_{\bm p,\bm q'} \left[(\Xi_c - \Xi_v)  F^*_{\bm p,\bm q'}(\bm k + \bm q) F_{\bm p,\bm q'}(\bm k) \right.\\
+\Xi_c F^*_{\bm p+\bm q,\bm q'}(\bm k + \bm q) F_{\bm p,\bm q'}(\bm k)\\
\left. 
- \Xi_c F^*_{\bm p,\bm q'+ \bm q}(\bm k + \bm q) F_{\bm p,\bm q'}(\bm k)\right].
\end{multline}
Under assumptions $k_F \ll q \ll a_{tr}^{-1}$ we can neglect the last term in Eq.~\eqref{V:ph:m} because the transition of a Fermi sea hole with large wavevector transfer is impossible, and summations over $\bm p$ and $\bm q$ can be reduced to $1$ by virtue of Eq.~\eqref{norm:chevy}. As a result, we arrive at
\[
M^{FP,\pm \bm q}_{\bm k\to \bm k'} = M^{tr,\pm \bm q}_{\bm k\to \bm k'},
\] 
where $M^{tr}$ is given by Eq.~\eqref{me} demonstrating the equivalence of the trion and Fermi-polaron approaches.

\end{document}